\documentclass[english]{IEEEtran}
\usepackage[T1]{fontenc}
\usepackage{color}
\usepackage{float}
\usepackage{units}
\usepackage{amstext}
\usepackage{graphicx}
\usepackage{esint}

\makeatletter
\usepackage{steinmetz}
\usepackage{hyperref}

\makeatother

\usepackage{babel}
\begin{document}

\title{Understanding early indicators of critical transitions in power systems
from autocorrelation functions}

\author{Goodarz Ghanavati, \emph{Student Member, IEEE}, Paul D.~H.~Hines,
\emph{Member, IEEE}, Taras I.~Lakoba, Eduardo Cotilla-Sanchez, \emph{Member,
IEEE}%
\thanks{This work was supported in part by the US Dept. of Energy, award \#DE-OE0000447,
 and in part by the US National Science Foundation, award \#ECCS-1254549
.

G. Ghanavati, P. Hines and T.Lakoba are with the College of Engineering
and Mathematical Sciences, University of Vermont, Burlington, VT (e-mail:
gghanava@uvm.edu; phines@uvm.edu; tlakoba@uvm.edu).

E. Cotilla-Sanchez is with the School of Electrical Engineering and
Computer Science at Oregon State University. (e-mail: cotillaj@eecs.oregonstate.edu).%
}}
\maketitle
\begin{abstract}
Many dynamical systems, including power systems, recover from perturbations
more slowly as they approach critical transitions---a phenomenon known
as critical slowing down. If the system is stochastically forced,
autocorrelation and variance in time-series data from the system often
increase before the transition, potentially providing an early warning
of coming danger. In some cases, these statistical patterns are sufficiently
strong, and occur sufficiently far from the transition, that they
can be used to predict the distance between the current operating
state and the critical point. In other cases CSD comes too late to
be a good indicator. In order to better understand the extent to which
CSD can be used as an indicator of proximity to bifurcation in power
systems, this paper derives autocorrelation functions for three small
power system models, using the stochastic differential algebraic equations
(SDAE) associated with each. The analytical results, along with numerical
results from a larger system, show that, although CSD does occur in
power systems, its signs sometimes appear only when the system is
very close to transition. On the other hand, the variance in voltage
magnitudes consistently shows up as a good early warning of voltage
collapse. Finally, analytical results illustrate the importance of
nonlinearity to the occurrence of CSD. \end{abstract}
\renewcommand\[{\begin{equation}}

\section{Introduction\label{sec:Introduction}}

There is increasing evidence that time-series data taken from stochastically
forced dynamical systems show statistical patterns that can be useful
in predicting the proximity of a system to critical transitions \cite{scheffer2009early},
\cite{lenton2012early}. Collectively this phenomenon is known as
Critical Slowing Down, and is most easily observed by testing for
autocorrelation and variance in time-series data. Increases in autocorrelation
and variance have been shown to give early warning of critical transitions
in climate models \cite{Slowclimate}, ecosystems \cite{ecosystemslow},
the human brain \cite{epilepticlitt2001} and electric power systems
\textcolor{black}{\cite{CSDjournal,Mitautocorrelation,podolsky2013critical}.}

Scheffer et al. \cite{scheffer2009early} provide some explanation
for why increasing variance and autocorrelation can indicate proximity
to a critical transition. They illustrate that increasing autocorrelation
results from the system returning to equilibrium more slowly after
perturbations, and that increased variance results from state variables
spending more time further away from equilibrium. Some further explanation
of CSD in stochastic systems can be found by looking at the theory
of fast-slow systems \cite{kuehn2011mathematical}. In many stochastic
systems with critical transitions there are two time scales; slow
trends gradually move the ``equilibrium'' operating state toward,
or away from points of instability, and random perturbations cause
fast changes in the state variables. In power systems, loads have
slow predictable trends, such as load ramps in the morning hours,
and fast stochastic ones, such as random load switching or rapid changes
in renewable generation. Reference \cite{kuehn2011mathematical} uses
the mathematical theory of the stochastic fast-slow dynamical systems
and the Fokker$ $--Planck equation to explain the use of autocorrelation
and variance as indicators of CSD.

While CSD is a general property of critical transitions~\cite{boerlijst2013catastrophic},
its signs do not always appear early enough to be useful as an early
warning, and do not universally appear in all variables~\cite{boerlijst2013catastrophic,hastings2010regime}.
References~\cite{boerlijst2013catastrophic} and \cite{hastings2010regime}
both show, using ecological models, that the signs of CSD appear only
in a few of the variables, or even not at all. 

Several types of critical transitions in deterministic power system
models have been explained using bifurcation theory. Reference \cite{Dobsonsaddle}
explains voltage collapse as a saddle-node bifurcation. Reference
\cite{canizares2002voltage} describes voltage instability caused
by the violation of equipment limits using limit-induced bifurcation
theory. Some types of oscillatory instability can be explained as
a Hopf bifurcation \cite{ajjarapu1992bifurcation}, \cite{Hopf_canizares}.
Reference \cite{CASequivalOPF} describes an optimization method that
can find saddle-node or limit-induced bifurcation points. Reference
\cite{CASmulti} shows that both Hopf and saddle-node bifurcations
can be identified in a multi-machine power system, and that their
locations can be affected by a power system stabilizer. In \cite{CASsingular},
authors computed the singular points of the differential and algebraic
equations that model the power system.

Substantial research has focused on estimating the proximity of a
power system to a particular critical transition. References \cite{canizares2002voltage},\cite{chiang1990voltage}\nocite{begovic1992control}--\cite{VanCutsem}
describe methods to measure the distance between an operating state
and voltage collapse with respect to slow-moving state variables,
such as load. Although these methods provide valuable information
about system stability, they are based on the assumption that the
current network model is accurate. However, all power system models
include error, both in state variable estimates and network parameters,
particularly for areas of the network that are outside of an operator's
immediate control.

An alternate approach to estimating proximity to bifurcation is to
study the response of a system to stochastic forcing, such as fluctuations
in load, or variable production from renewable energy sources. To
this end, a growing number of papers study power system stability
using stochastic models \cite{Bergen1987}\nocite{nwankpa1992stochastic,anghel2007stochastic,Hillstochastic}--\cite{wangfokker}.
Reference \cite{Bergen1987} models power systems using Stochastic
Differential Equations (SDEs) in order to develop a measure of voltage
security. In \cite{Hillstochastic}, numerical methods are used to
assess transient stability in power systems, given fluctuating loads
and random faults. Reference \cite{wangfokker} uses the Fokker-Planck
equation to calculate the probability density function (PDF) for state
variables in a single machine infinite bus system (SMIB), and uses
the time evolution of this PDF to show how random load fluctuations
affect system stability. 

The results above clearly show that power system stability is affected
by stochastic forcing. However, they provide little information about
the extent to which CSD can be used as an early warning of critical
transitions given fluctuating measurement data. Given the increasing
availability of high-sample-rate synchronized phasor measurement unit
(PMU) data, and the fact that insufficient situational awareness has
been identified as a critical contributor to recent large power system
failures (e.g., \cite{August2003}, \cite{Southwest}) there is a
need to better understand how statistical phenomena, such as CSD,
might be used to design good indicators of stress in power systems.

Results from the literature on CSD suggest that autocorrelation and
variance in time-series data increase before critical transitions.
Empirical evidence for increasing autocorrelation and variance is
provided for an SMIB and a 9-bus test case in \textcolor{black}{\cite{CSDjournal}.
}Reference \cite{ChertkovCSD} shows that voltage variance at the
end of a distribution feeder increases as it approaches voltage collapse.
However, the results do not provide insight into autocorrelation.
To our knowledge, only \cite{Mitautocorrelation,podolsky2013critical}
derive approximate analytical autocorrelation functions (from which
either autocorrelation or variance can be found) for state variables
in a power system model, which is applied to the New England 39 bus
test case. However, the autocorrelation function in \cite{Mitautocorrelation,podolsky2013critical}
is limited to the operating regime very close to the threshold of
system instability. Furthermore, there is, to our knowledge, no existing
research regarding which variables show the signs of CSD most clearly
in power system, and thus which variables are better indicators of
proximity to critical transitions. In \cite{Ghanavati:2013}, the authors derived 
the general autocorrelation function for the stochastic SMIB system. 
This paper extends the SMIB results in \cite{Ghanavati:2013}, and studies two additional 
power system models using the same analytical approach. Also, this paper 
includes new numerical simulation results for two multi-machine systems,
 which illustrate insights gained from the analytical work.

Motivated by the need to better understand CSD in power systems, the
goal of this paper is to describe and explain changes in the autocorrelation
and variance of state variables in several power system models, as
they approach bifurcation\textcolor{black}{. To this end, we derive
autocorrelation functions of state variables for three small models.
We use} the results to show that CSD does occur in power systems,
explain why it occurs, and describe conditions under which autocorrelation
and variance signal proximity to critical transitions. The remainder
of this paper is organized as follows. Section \ref{sec:SDE} describes
the general mathematical model and the method used to derive autocorrelation
functions in this paper. Analytical solutions and illustrative numerical
results for three small power systems are presented in Secs.~\ref{sec:SMIB},
\ref{sec:voltage-collapse} and \ref{sec:Three-Bus-System}. In Sec.
\ref{sec:New-England-39-Bus}, the results of numerical simulations
on two multi-machine power system models including the New England
39 bus test case are presented. Finally, Sec.~\ref{sec:Conclusion}
summarizes the results and contributions of this paper.

\section{Solution Method for Autocorrelation Functions\label{sec:SDE}}

In this section, we present the general form of the Stochastic Differential
Algebraic Equations (SDAEs) used to model the three systems studied
in this paper. Then, the solution of the SDAEs and the expressions
for autocorrelations and variances of both algebraic and differential
variables of the systems are presented. Finally, the method used for
simulating the SDAEs numerically is described.

\vspace{-.2in}
\subsection{The Model\label{sub:SDE-Model}}

All three models studied analytically in this paper include a single
second-order synchronous generator. These systems can be described
by the following SDAEs:
\begin{eqnarray}
\ddot{\delta}+2\gamma\dot{\delta}+F_{1}\left(\delta,\underline{y},\eta\right) & = & 0\label{eq:2-1}\\
\underline{F_{2}}\left(\delta,\underline{y},\eta\right) & = & 0\label{eq:2-2}
\end{eqnarray}
where $\delta$ is angle of the synchronous generator's rotor relative
to a synchronously rotating reference axis, $\underline{y}$ is the
vector of algebraic variables, $\gamma$ is the damping coefficient,
$F_{1},\underline{F_{2}}$ form a set of nonlinear algebraic equations
of the systems, and $\eta$ is a Gaussian random variable. $\eta$
has the following properties:
\begin{eqnarray}
\textnormal{{E}}\left[\eta\left(t\right)\right] & = & 0\label{eq:2-3}\\
\textnormal{{E}}\left[\eta\left(t\right)\eta\left(s\right)\right] & = & \sigma_{\eta}^{2}\cdot\delta_{I}\left(t-s\right)\label{eq:2-4}
\end{eqnarray}
where $t,s$ are two arbitrary times, $\sigma_{\eta}^{2}$ is the
intensity of noise, and $\delta_{I}$ represents the unit impulse
(delta) function (which should not be confused with the rotor angle
$\delta$). There are a variety of sources of noise, such as random
load switching or variable renewable generation,  in power systems.
To our knowledge, no existing studies have  quantified the correlation
time of noise in power systems. Thus, in this paper, we assume that
the correlation time of noise is negligible relative to the response-time
of the system, which means that $\mbox{E}[\eta(t)\eta(s)]=0$ for
all $s$ significantly greater than $t$. It is important to note
that the variance of $\eta$ is infinite according to~(\ref{eq:2-4}),
because the delta function is infinite at $t=s$, which means that
particular care is needed when simulating~(\ref{eq:2-1}) and~(\ref{eq:2-2})
numerically (see Sec.~\ref{sub:Ar-and-Var-algebraic}).

In order to solve~(\ref{eq:2-1}) and~(\ref{eq:2-2}) analytically,
we linearized $F_{1}$ and $\underline{F_{2}}$ around the stable
equilibrium point using first-order Taylor expansion. Then~(\ref{eq:2-1})
and~(\ref{eq:2-2}) were combined into a single damped harmonic oscillator
equation with stochastic forcing:
\begin{equation}
\Delta\ddot{\delta}+2\gamma\Delta\dot{\delta}+\omega_{0}^{2}\Delta\delta=-f\eta\label{eq:2-5}
\end{equation}
where $\omega_{0}$ is the undamped angular frequency of the oscillator,
$f$ is a constant, and $\Delta\delta=\delta-\delta_{0}$ is the deviation
of the rotor angle from its equilibrium value. Both $\omega_{0}$
and $f$ change with the system's equilibrium operating state. Equation
(\ref{eq:2-5}) can be written as a multivariate Ornstein--Uhlenbeck
process \cite{Gardiner:2004}:
\begin{equation}
\dot{\underline{z}}\left(t\right)=A\underline{z}\left(t\right)+B\left[\begin{array}{c}
0\\
\eta\left(t\right)
\end{array}\right]\label{eq:1-1}
\end{equation}
where $\underline{z}=\left[\begin{array}{cc}
\Delta\delta & \Delta\dot{\delta}\end{array}\right]^{T}$ is the vector of differential variables, $\Delta\dot{\delta}$ is
the deviation of the generator speed from its equilibrium value, and
$A$ and $B$ are constant matrices as follows:
\begin{eqnarray}
A & = & \left[\begin{array}{cc}
0 & 1\\
-\omega_{0}^{2} & -2\gamma
\end{array}\right]\label{eq:4-1}\\
B & = & \left[\begin{array}{cc}
0 & 0\\
0 & -f
\end{array}\right]\label{eq:4-2}
\end{eqnarray}
Given (\ref{eq:4-1}), the eigenvalues of $A$ are $-\gamma\pm\sqrt{\gamma^{2}-\omega_{0}^{2}}$.
At $\omega_{0}=0$, one of the eigenvalues of matrix $A$ becomes
zero, and the system experiences a saddle-node bifurcation. 

Equation (\ref{eq:2-5}) can be interpreted in two different ways:
using either $\mathbf{\textrm{It\ensuremath{\hat{o}}}}$ SDE and Stratonovich
SDEs. In the $\mathbf{\textrm{It\ensuremath{\hat{o}}}}$ interpretation~\cite{ItoSDE},
noise is considered to be uncorrelated. However, in the Stratonovich
interpretation \cite{Stratonovich:1963}, which is a more natural
choice physically, noise has finite, albeit very small, correlation
time \cite{Gardiner:2004}. $\mathbf{\textrm{It\ensuremath{\hat{o}}}}$
calculus is often used in discrete systems, such as finance, though
a few papers have applied the $\mathbf{\textrm{It\ensuremath{\hat{o}}}}$
approach to power systems \cite{Bergen1987}, \cite{Hillstochastic}.
On the other hand, the Stratonovich method is often used in continuous
physical systems or systems with band-limited noise \cite{mannella2012ito}.
The Stratonovich interpretation also allows the use of ordinary calculus,
which is not possible with the $\mathbf{\textrm{It\ensuremath{\hat{o}}}}$
interpretation. Because $B$ is a constant matrix in this paper, the
$\mathbf{\textrm{It\ensuremath{\hat{o}}}}$ and Stratonovich interpretations
result in the same solution~\cite{mannella2012ito}. This paper follows
the Stratonovich interpretation because it allows one to use ordinary
calculus. 

Following the method in \cite{Diff.equations}, if $\gamma<\omega_{0}$
(which holds until very close to the bifurcation in two of our systems),
the solution of (\ref{eq:1-1}) is as follows:
\begin{eqnarray}
\Delta\delta(t) & = & f\cdot\int_{-\infty}^{t}\exp\left(\gamma\left(t'-t\right)\right)\eta\left(t'\right)\cdot\label{eq:22-2-1}\\
 &  & \qquad\frac{\sin\left(\omega'(t'-t)\right)}{\omega'}dt'\nonumber \\
\Delta\dot{\delta}\left(t\right) & =- & f\cdot\int_{-\infty}^{t}\exp\left(\gamma\left(t'-t\right)\right)\eta\left(t'\right)\cdot\label{eq:22-3-1}\\
 &  & \qquad\frac{\sin\left(\omega'(t'-t)+\phi\right)\omega_{0}}{\omega'}dt'\nonumber 
\end{eqnarray}
where $t'$ is the variable of integration, $\omega'=\sqrt{\omega_{0}^{2}-\gamma^{2}}$
is the frequency of the underdamped harmonic oscillator, and $\phi=\arctan(\omega'/\gamma)$.

In the system considered in Sec.~\ref{sec:voltage-collapse}, $\omega_{0}$
in (\ref{eq:2-5}) is equal to zero for all system parameters, so
the condition $\gamma<\omega_{0}$ does not hold. Therefore, the solution
of (\ref{eq:2-5}) in that system is different from (\ref{eq:22-2-1}),
(\ref{eq:22-3-1}) as follows:
\begin{equation}
\Delta\dot{\delta}=-f\int_{-\infty}^{t}\exp\left(-2\gamma\left(t-t'\right)\right)\eta\left(t'\right)dt'\label{eq:37-1}
\end{equation}

\subsection{Autocorrelation and Variance of Differential Variables\label{sub:Calculation-of-Autocorrelations}}

Given that the eigenvalues of $A$ have negative real part before
the bifurcation (because $\gamma>0$), one can calculate the stationary
variances and autocorrelations of $\Delta\delta$ and $\Delta\dot{\delta}$
using (\ref{eq:2-3}), (\ref{eq:2-4}), (\ref{eq:22-2-1}) and (\ref{eq:22-3-1}).
The variances of the differential variables are as follows:
\begin{eqnarray}
\sigma_{\Delta\delta}^{2} & = & \frac{f^{2}\sigma_{\eta}^{2}}{4\gamma\omega_{0}^{2}}\label{eq:8-1-1}\\
\sigma_{\Delta\dot{\delta}}^{2} & = & \frac{f^{2}\sigma_{\eta}^{2}}{4\gamma}\label{eq:8-2-1}
\end{eqnarray}
If $\gamma<\omega_{0}$, the normalized autocorrelation functions
for $\Delta\delta$ and $\Delta\dot{\delta}$ are as follows:
\begin{eqnarray}
\frac{\text{E}\left[\Delta\delta\left(t\right)\Delta\delta\left(s\right)\right]}{\sigma_{\Delta\delta}^{2}} & = & \exp\left(-\gamma\Delta t\right)\frac{\omega_{0}}{\omega'}\cdot\label{eq:10-1-1}\\
 &  & \sin\left(\omega'\Delta t+\phi\right)\nonumber \\
\frac{\text{E}\left[\Delta\dot{\delta}\left(t\right)\Delta\dot{\delta}\left(s\right)\right]}{\sigma_{\Delta\dot{\delta}}^{2}} & = & \exp\left(-\gamma\Delta t\right)\frac{-\omega_{0}}{\omega'}\cdot\label{eq:10-2-1}\\
 &  & \sin\left(\omega'\Delta t-\phi\right)\nonumber 
\end{eqnarray}
where $\Delta t=t-s$.

If $\omega_{0}=0$, the variance of $\Delta\dot{\delta}$ can be calculated
from (\ref{eq:8-2-1}) and the autocorrelation of $\Delta\dot{\delta}$
is as follows:

\begin{equation}
\frac{\text{E}\left[\Delta\dot{\delta}\left(t\right)\Delta\dot{\delta}\left(s\right)\right]}{\sigma_{\Delta\dot{\delta}}^{2}}=\exp\left(-2\gamma\Delta t\right)\label{eq:38-2}
\end{equation}

\subsection{Autocorrelation and Variance of Algebraic Variables\label{sub:Ar-and-Var-algebraic}}

\textcolor{black}{In order to compute} the autocorrelation functions
of the algebraic variables, we calculated the algebraic variables
as linear functions of the differential variable $\Delta\delta$ and
the noise $\eta$, by linearizing $\underline{F_{2}}$ in (\ref{eq:2-2}):
\begin{equation}
\Delta y_{i}\left(t\right)=C_{i_{,}1}\Delta\delta\left(t\right)+C_{i_{,}2}\eta\label{eq:14}
\end{equation}
where $y_{i}$ is an algebraic variable, and $C_{i_{,}1},C_{i_{,}2}$
are constant values. Then, the autocorrelation of $\Delta y_{i}$
is as follows for $t\geq s$:
\begin{eqnarray}
\text{E}\left[\Delta y_{i}\left(t\right)\Delta y_{i}\left(s\right)\right] & = & C_{i_{,}1}^{2}\cdot\text{E}\left[\Delta\delta\left(t\right)\Delta\delta\left(s\right)\right]+\label{eq:25-1}\\
 &  & C_{i_{,}1}C_{i_{,}2}\mbox{\ensuremath{\cdot}}\textnormal{E}\left[\Delta\delta\left(t\right)\eta\left(s\right)\right]+\nonumber \\
 &  & C_{i_{,}2}^{2}\cdot\text{E}\left[\eta\left(t\right)\eta\left(s\right)\right]\nonumber 
\end{eqnarray}
In deriving (\ref{eq:25-1}), we used the fact that $\textnormal{{E}}\left[\Delta\delta\left(s\right)\eta\left(t\right)\right]=0$
since the system is causal. Equation (\ref{eq:25-1}) shows that,
in order to calculate the autocorrelation of $\Delta y_{i}\left(t\right)$,
it is necessary to calculate $\textnormal{{E}}\left[\Delta\delta\left(t\right)\eta\left(s\right)\right]$.
Using~(\ref{eq:22-2-1}), $\textnormal{{E}}\left[\Delta\delta\left(t\right)\eta\left(s\right)\right]$
is as follows:
\begin{eqnarray}
\textnormal{{E}}\left[\Delta\delta\left(t\right)\eta\left(s\right)\right] & = & -\exp\left(-\gamma\Delta t\right)\cdot\frac{f}{\omega'}\cdot\label{eq:24-1}\\
 &  & \sin\left(\omega'\Delta t\right)\sigma_{\eta}^{2}\nonumber 
\end{eqnarray}
which indicates that  $\textrm{cov}\left(\Delta\delta,\eta\right)=0$.

In order to use (\ref{eq:25-1}) to compute the variance of $\Delta y_{i}$,
we need to carefully consider our model of noise in numerical computations.
According to (\ref{eq:2-4}), the variance of $\eta$ is infinite,
because the delta function is infinite at $t=s$, which would mean
that the variance of $\Delta y_{i}$ could be infinite. However, the
noise in numerical simulations must have a finite variance. To determine
it, we rewrite (\ref{eq:1-1}) as follows:
\begin{equation}
d\underline{z}\left(t\right)=A\underline{z}\left(t\right)dt+Bd\underline{W}\left(t\right)\label{eq:wiener}
\end{equation}
where $d\underline{W}\left(t\right)=\eta dt$ is the Wiener process.
It is well-known that the variance of $dW\left(t\right)$ is $\sigma_{\eta}^{2}dt$
\cite{Gardiner:2004}. In numerical simulations, $dt=\tau_{\textnormal{{int}}}$,
where $\tau_{\textnormal{{int}}}$ is the integration time step. Thus,
${E}\left[dW_{\textnormal{{num}}}^{2}\right]=\textnormal{{E}}\left[\left(\eta_{\textnormal{{num}}}\tau_{{int}}\right)^{2}\right]=\sigma_{\eta}^{2}\tau_{\textnormal{{int}}}$.
Hence, ${E}\left[\eta{}_{\textnormal{{num}}}^{2}\right]=\sigma_{\eta}^{2}/\tau_{\textnormal{{int}}}$.
With this definition of noise,  (\ref{eq:14}) means that the variance
of $\Delta y_{i}$ is:

\begin{equation}
\sigma_{\Delta y_{i}}^{2}=C_{i_{_{,}}1}^{2}\sigma_{\Delta\delta}^{2}+C_{i_{,}2}^{2}\frac{\sigma_{\eta}^{2}}{\tau_{\textrm{int}}}\label{eq:24-3}
\end{equation}
where $\tau_{\textrm{int}}$ is the integration time step in numerical
simulations.  In order to match analytical results with numerical
simulations, we divided the noise intensity by the integration step
size in the second term of the right-hand side of (\ref{eq:24-3}).
Combining (\ref{eq:8-1-1}) and (\ref{eq:24-3}) results in the following:

\begin{equation}
\sigma_{\Delta y_{i}}^{2}=\left(\frac{C_{i_{,}1}^{2}f^{2}}{4\gamma\omega_{0}^{2}}+\frac{C_{i_{,}2}^{2}}{\tau_{\textrm{int}}}\right)\sigma_{\eta}^{2}\label{eq:24-2}
\end{equation}
Combining (\ref{eq:8-1-1}), (\ref{eq:10-1-1}), (\ref{eq:25-1})
, (\ref{eq:24-1}) and (\ref{eq:24-2}), we calculated the normalized
autocorrelation function of $\Delta y_{i}$:

\begin{eqnarray}
\frac{\textnormal{{E}}\left[\Delta y_{i}\left(t\right)\Delta y_{i}\left(s\right)\right]}{\sigma_{\Delta y_{i}}^{2}} & = & \exp\left(-\gamma\Delta t\right)\sin\left(\omega'\Delta t+\phi_{\Delta y_{i}}\right)\cdot\nonumber \\
 &  & \frac{C_{i_{,}1}f\omega_{0}\sqrt{\lambda}}{\omega'\left(C_{i_{,}1}^{2}f^{2}+4C_{i_{,}2}^{2}\gamma\omega_{0}^{2}\right)}\label{eq:25-3}
\end{eqnarray}
where $\lambda=\sqrt{C_{i_{,}1}f\left(C_{i_{,}1}f-8C_{i_{,}2}\gamma^{2}\right)+\left(4C_{i_{,}2}\omega_{0}\gamma\right)^{2}},\phi_{\Delta y_{i}}=\arctan\left(\nicefrac{C_{i_{,}1}f\omega'}{\left(C_{i_{,}1}f\gamma-4C_{i_{,}2}\gamma\omega_{0}^{2}\right)}\right)$.

\subsection{Numerical Simulation\label{sub:Numerical-Simulation}}

In order to calculate numerical results that can be compared to the
analytical ones, (\ref{eq:2-1}) and (\ref{eq:2-2}) were solved using
a trapezoidal ordinary differential equation solver, with a fixed
time step of integration, $\tau_{\textrm{int}}$. We chose the integration
step size $\tau_{\textrm{int}}$ to be much shorter than the the smallest
period of oscillation $T=2\pi/\omega'$, between the periods for all
bifurcation parameter values. 

In order to determine numerical mean values in this paper, each set
of SDEs was simulated 100 times. In each case the resulting averages
were compared with analytical means.

\section{Single Machine Infinite Bus System\label{sec:SMIB}}

Analysis of small power system models can be very helpful for understanding
the concepts of power system stability. The single machine infinite
bus system has long been used to understand the behavior of a relatively
small generator connected to a larger system through a long transmission
line. This SMIB system has been used, for example, to  explore the
small signal stability of synchronous machines \cite{demello1969concepts}
and to evaluate control techniques to improve transient stability
and voltage regulation \cite{wangHill1993SMIB}.  In the recent literature,
there is increasing interest in stochastic behavior of power systems,
in part due to the increasing integration of variable renewable energy
sources. A few of these papers use stochastic SMIB models. In \cite{wei2009SMIB},
it is suggested that increasing noise in the stochastic SMIB system
can make the system unstable and induce chaotic behavior. Reference
\cite{wangfokker} (mentioned in Sec.~\ref{sec:Introduction}) also
studied stability in a stochastic SMIB system.

In this section, we use the autocorrelation functions derived in Sec.~\ref{sec:SDE}
to calculate the variances and autocorrelations of the state variables
of a stochastic SMIB system. Analysis of these functions provides
analytical evidence for, and insight into, CSD in a small power system.

\vspace{-.2in}
\subsection{Stochastic SMIB System Model\label{sec: SMIB-1}}

Fig. \ref{SMIB} shows the stochastic SMIB system. Equation (\ref{eq:1}),
which combines the mechanical swing equation and the electrical power
produced by the generator, fully describes the dynamics of this system:
\begin{equation}
M\ddot{\delta}+D\dot{\delta}+\frac{(1+\eta)E'_{a}}{X}\sin\left(\delta\right)=P_{m}\label{eq:1}
\end{equation}
where $(\eta\sim\mathcal{N}(0,0.01))$ is a white Gaussian random
variable added to the voltage magnitude of the infinite bus to account
for the noise in the system, $M$ and $D$ are the combined inertia
constant and damping coefficient of the generator and turbine, and
$E_{a}^{'}$ is the transient emf. The reactance $X$ is the sum of
the generator transient reactance $(X_{d}^{'})$ and the line reactance
$\left(X_{l}\right)$, and $P_{m}$ is the input mechanical power.
The value of parameters used in this section are given below: 

\begin{center}
$D=0.03\frac{\textrm{pu}}{rad/s},H=4\frac{MW.s}{MVA},X_{d}'=0.15\textrm{pu,}$
\par\end{center}

\begin{center}
$X_{l}=0.2\textrm{pu},\omega_{s}=2\pi\cdot60\textrm{{rad/s}}$
\par\end{center}

\noindent Note that $M=2H/\omega_{s}$, where $H$ is the inertia
constant in seconds, and $\omega_{s}$ is the rated speed of the machine.
The generator and the system base voltage levels are $13.8\text{kV}$
and $115\textrm{{kV}}$, and both the generator and system per unit
base are set to $100$MVA. The generator transient reactance $X_{d}^{'}=0.15\cdot(13.8/115)^{2}$
pu, on the system pu base. The third term on the left-hand side of
(\ref{eq:1}) is the generator's electrical power $\left(P_{g}\right)$.

In order to test the system at various load levels, we solved the
system for different equilibria, with the generator's mechanical and
electrical power equal at each equilibrium:
\begin{equation}
P_{m}=P_{g0}=\frac{E_{a}^{'}}{X}\sin\left(\delta_{0}\right)\label{eq:1-2}
\end{equation}
where $\delta_{0}$ is the value of the generator rotor angle at equilibrium. 

\begin{figure}[H]
\begin{centering}
\includegraphics[width=1\columnwidth]{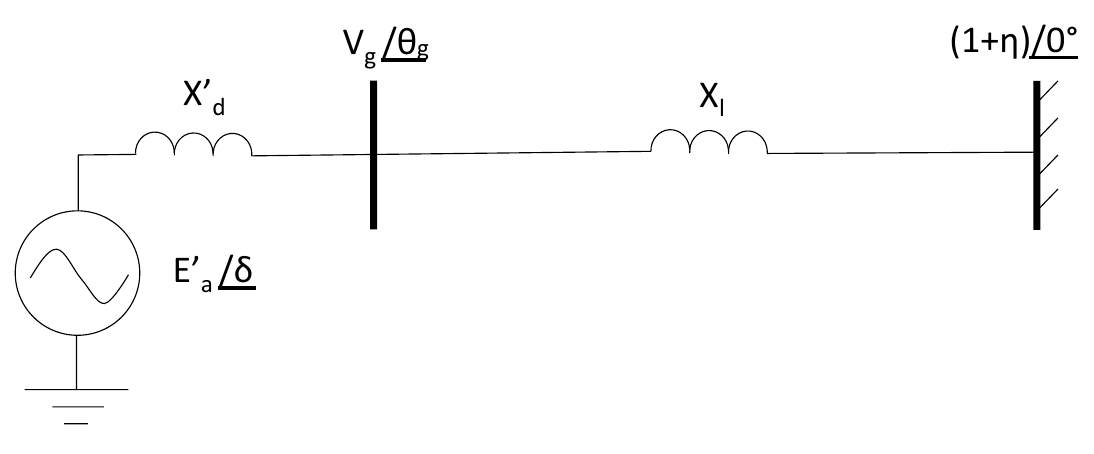}
\par\end{centering}

\caption{\label{SMIB}Stochastic single machine infinite bus system used in
Sec. \ref{sec:SMIB}. The notation $V_{g}\phase{\theta_{g}}$ represents
$V_{g}\exp\left[j\theta_{g}\right]$. }
\end{figure}

\subsection{Autocorrelation and Variance \label{sec: SMIB-2}}

In this section, we calculate the autocorrelations and variances of
the algebraic and differential variables of this system using the
method in Sec.~\ref{sec:SDE}. Equations (\ref{eq:2-1}) and (\ref{eq:2-2})
describe this system for which the following equalities hold:
\begin{equation}
\gamma=\frac{D}{2M};\omega_{0}=\sqrt{\frac{E_{a}^{'}\cos\delta_{0}}{MX}};\underline{y}=\left[\begin{array}{cc}
V_{g} & \theta_{g}\end{array}\right]^{T}\label{eq:18}
\end{equation}
\begin{equation}
f=\frac{P_{g0}}{M};F_{1}\left(z,\underline{y},\eta\right)=\left(\frac{(1+\eta)E_{a}^{'}}{X}\sin\delta-P_{m}\right)/M\label{eq:19}
\end{equation}
where $\Delta V_{g}=V_{g}-V_{g0},\Delta\theta_{g}=\theta_{g}-\theta_{g0}$
are the deviations of, respectively the generator terminal busbar's
voltage magnitude and angle from their equilibrium values. Equations
(\ref{eq:18}) and (\ref{eq:19}) show that $f$ increases with $\delta_{0}$
while $\omega_{0}$ decreases with $\delta_{0}$. 

In order to calculate the algebraic equations, which form $\underline{F_{2}}\left(\delta,y,\eta\right)$
in (\ref{eq:2-2}), we wrote Kirchhoff's current law at the generator's
terminal:
\begin{equation}
\frac{E'_{a}e^{j\delta}-V_{g}e^{j\theta_{g}}}{jX'_{d}}+\frac{1+\eta-V_{g}e^{j\theta_{g}}}{jX_{l}}=0\label{eq:11-1}
\end{equation}
Separating the real and imaginary parts in (\ref{eq:11-1}) gives
the following:
\begin{eqnarray}
V_{g}\sin\left(\theta_{g}\right) & = & \alpha E_{a}^{'}\sin\left(\delta\right)\label{eq:11-2}\\
V_{g}\cos\left(\theta_{g}\right) & = & \alpha E_{a}^{'}\cos\left(\delta\right)\label{eq:11-3}\\
 &  & +\left(1+\eta\right)\left(1-\alpha\right)\nonumber 
\end{eqnarray}
where $\alpha=X_{l}/(X_{l}+X_{d}^{'})$. Equations (\ref{eq:11-2})
and (\ref{eq:11-3}) combine to make $\underline{F_{2}}\left(\delta,y,\eta\right)$
in (\ref{eq:2-2}). 

Linearizing (\ref{eq:11-2}) and (\ref{eq:11-3}) yields the coefficients
in (\ref{eq:14}), which are necessary for calculating the autocorrelations
and variances of the algebraic variables (here $y_{1}=\Delta V_{g},y_{2}=\Delta\theta_{g}$):
\begin{eqnarray}
C_{1_{,}1} & = & \alpha E_{a}^{'}\sin\left(\theta_{g0}-\delta_{0}\right)\label{eq:11-4}\\
C_{1_{,}2} & = & \left(1-\alpha\right)\cos\left(\theta_{g0}\right)\label{eq:11-5}\\
C_{2_{,}1} & = & \alpha E_{a}^{'}\cos\left(\theta_{g0}-\delta_{0}\right)\label{eq:11-6}\\
C_{2_{,}2} & = & -\left(1-\alpha\right)\sin\left(\theta_{g0}\right)\label{eq:11-7}
\end{eqnarray}

Fig. \ref{fig:omega0_f} shows the decrease of $\omega'$, which is
the absolute value of the imaginary part of the eigenvalues of $A$
in (\ref{eq:4-1}), with $P_{m}$. Note that the bifurcation occurs
at $P_{m}=5\textrm{{pu}}$. This figure illustrates how it can be
difficult to accurately foresee a bifurcation  by computing the eigenvalues
of a system (as in, e.g., \cite{chiang1990voltage}), if there is
noise in the measurements feeding the calculation.  The value of $\omega'\sim\left(P_{m}-\textrm{{b}}\right)^{1/4}$
does not decrease by a factor of two (compared to  its value at $P_{m}=1.0\textnormal{{pu}}$)
until $P_{m}=4.83\textrm{{pu}}$ (only $<3.4\%$ away from the bifurcation).
It decreases by another factor of two at $P_{m}=4.99\textrm{{pu}}$
($0.2\%$ away from the bifurcation). Also, note that the real part
of the eigenvalues are equal to $-\gamma$ until very close to the
bifurcation ($0.1\%$ away from the bifurcation), so they do not provide
a useful indication of proximity to the bifurcation. Thus, one can
confidently predict from $\omega'$ the imminent occurrence of the
bifurcation only very near it, which may be too late to avert it.
On the other hand, we will demonstrate below that for this system,
autocorrelation functions can provide substantially more advanced
warning of the bifurcation.

\begin{figure}
\includegraphics[width=1\columnwidth]{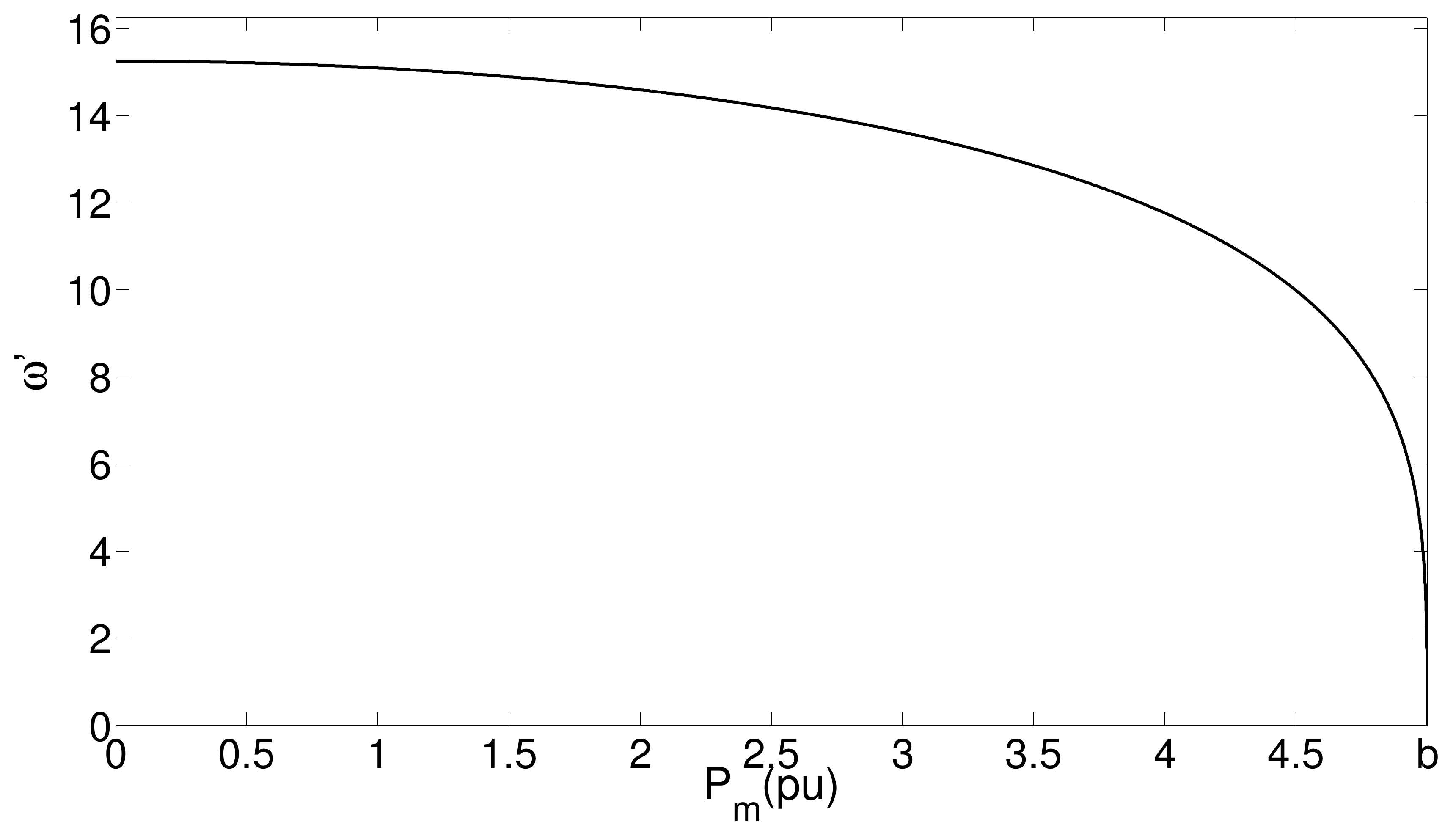}
\vspace{-.2in}
\caption{\label{fig:omega0_f}The decrease of $\omega'$ with $P_{m}$ in the
SMIB system. Near the bifurcation, $\omega'$ is very sensitive to
changes in $P_{m}$. In this figure, and most that follow, b is the
value of the bifurcation parameter ($P_{m}$ in this system) at the
bifurcation.}
\end{figure}

Using autocorrelation  as an early warning sign of potential bifurcations
requires that one carefully select a time lag, $\Delta t=t-s$, such
that changes in autocorrelation are observable. To understand the
impact of different time lags, we computed the autocorrelation as
function of $\Delta\delta$ (see Fig. \ref{fig: A_fcn}). From (\ref{eq:10-1-1}),
the autocorrelation of $\delta(t)$ crosses zero at  $\Delta t_{0}=\frac{2\pi-\phi}{\omega'}$.
The implication is that choosing $\Delta t$ close to $\Delta t_{0}$
allows one to observe a monotonic increase of autocorrelation as $P_{m}$
increases. For $\Delta t>\Delta t_{0}$, autocorrelation may not increase
monotonically, or the autocorrelation for some values of $P_{m}$
may be negative. For example, in Fig. \ref{fig: A_fcn} for $\Delta t=0.3\textrm{{s}}$,
the autocorrelation decreases first and then increases with $P_{m}$.
On the other hand, for $\Delta t$ considerably smaller than $\Delta t_{0}$,
the increase of the autocorrelation may not be large enough to be
measurable. In Fig. \ref{fig: A_fcn}, the curves converge as $\Delta t\rightarrow0$.
\textcolor{black}{Given that the smallest period of oscillation $(T=2\pi/\omega')$
in this system is $0.41\textrm{s}$,  we chose $\Delta t=0.1$s for
the autocorrelation calculations in this section. }

\begin{figure}
\begin{centering}
\includegraphics[width=1\columnwidth]{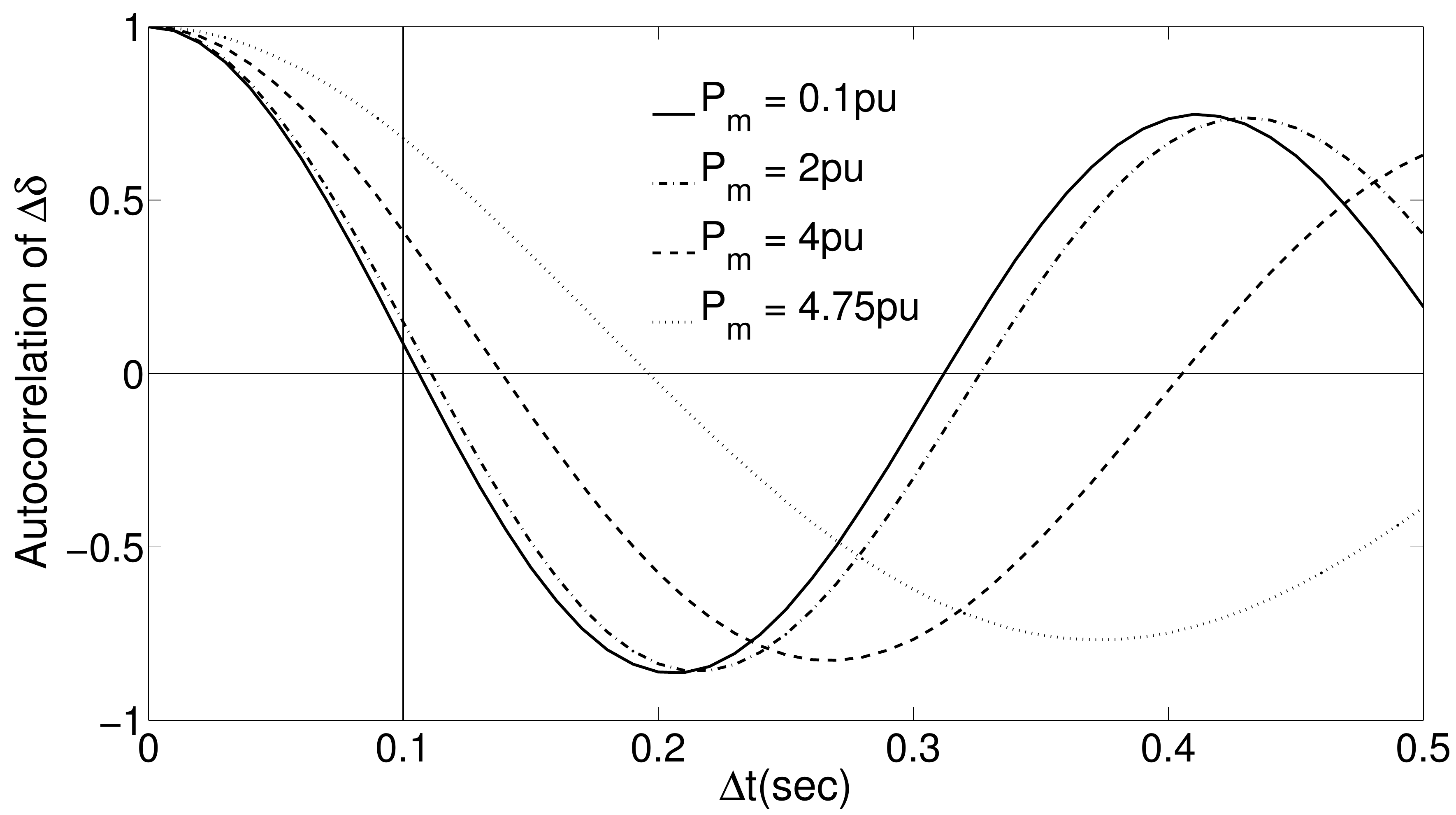}
\par\end{centering}
\vspace{-.1in}
\caption{\label{fig: A_fcn}Autocorrelation function of $\Delta\delta$. $\Delta t=0.1$s
is close to 1/4 of the smallest period of the function for all values
of $P_{m}$.}
\vspace{-.1in}
\end{figure}

Using (\ref{eq:8-1-1})--(\ref{eq:10-2-1}), we calculated the variances
and autocorrelations of $\Delta\delta$, $\Delta\dot{\delta}$ at
different operating points. In Figs.~\ref{fig_sys1_var_deom} and~\ref{fig_sys1_ar_deom},
these analytical results are compared with the numerical ones. To
initialize the numerical simulations, we assumed that $V_{g0}=1\textrm{{pu}}$
and solved for $E'_{a}$ in (\ref{eq:11-2}), (\ref{eq:11-3}) to
obtain $V_{g}=V_{g0}$ (for $\eta=0$). We chose the integration step
size $\tau_{int}$ to be 0.01s, which is much shorter than the the
smallest period of oscillation ($T=0.41\textrm{s}$). The numerical
results are shown for the range of  bifurcation parameter values for
which the numerical solutions were stable. 

In order to determine if variance and autocorrelation measurably increase
as load approaches the bifurcation, we computed the ratio of each
statistic when load is at 80\% of the bifurcation value to the value
when load is at 20\% of b. This ratio, $q_{\frac{80}{20}}$ in Figs.~\ref{fig_sys1_var_deom}
and~\ref{fig_sys1_ar_deom}, is defined as follows:
\begin{equation}
q_{\frac{80}{20}}=\frac{\textrm{Autocorrelation}\textrm{ {of} }u\textrm{ or }\sigma_{u}^{2}|_{P_{m}=0.8b}}{\textrm{Autocorrelation}\textrm{ {of} }u\textrm{ or }\sigma_{u}^{2}|_{P_{m}=0.2b}}\label{eq:10-3}
\end{equation}
where $u$ is the plot's variable. In subsequent figures, $q_{\frac{80}{20}}$
is defined similarly.

Fig.~\ref{fig_sys1_var_deom} shows that the variances of both $\Delta\delta$
and $\Delta\dot{\delta}$ increase substantially with $P_{m}$, and
thus appear to be good warning signs of the bifurcation. However,
the two variances grow with different rates. (This becomes clear when
comparing the ratios $q_{\frac{80}{20}}$ for $\Delta\delta$ and
$\Delta\dot{\delta}$.) The difference becomes even more noticeable
near the bifurcation where the variance of $\Delta\delta$ increases
much faster than the variance of $\Delta\dot{\delta}$. This is caused
by the term $\omega_{0}^{2}$ in the denominator of the expression
for the variance of $\Delta\delta$ in (\ref{eq:8-1-1}). In Fig.~\ref{fig_sys1_ar_deom},
the autocorrelations of $\Delta\delta$ and $\Delta\dot{\delta}$
increase with $P_{m}$. Similar to the variances, the autocorrelations
are good early warning signs of the bifurcation as well. Comparing
Figs.~\ref{fig_sys1_var_deom} and \ref{fig_sys1_ar_deom} with Fig.~\ref{fig:omega0_f}
(where an equivalent $q_{\frac{80}{20}}$ would be 1.28) shows that
the autocorrelations and variances of $\Delta\delta$ and $\Delta\dot{\delta}$
provide a substantially stronger early warning sign, relative to using
eigenvalues to estimate the distance to bifurcation in this system. 

\begin{figure}
\begin{centering}
\includegraphics[width=1\columnwidth,height=0.2\textheight]{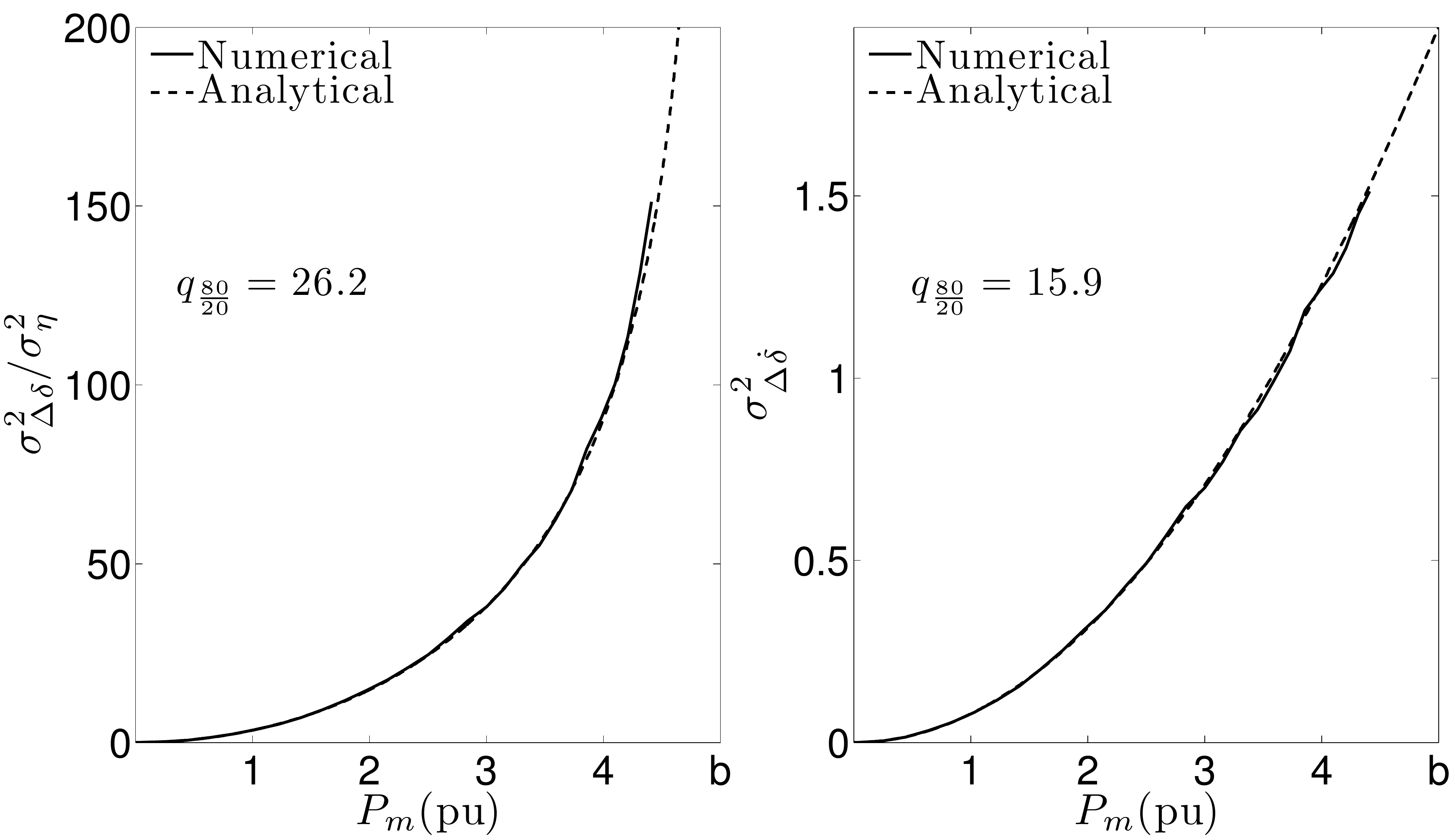}
\par\end{centering}
\vspace{-.15in}
\caption{\label{fig_sys1_var_deom}Variances of $\Delta\delta,\Delta\dot{\delta}$
versus mechanical power $(P_{m})$ values.}
\vspace{-.1in}
\end{figure}

\begin{figure}
\begin{centering}
\includegraphics[width=1\columnwidth,height=0.2\textheight]{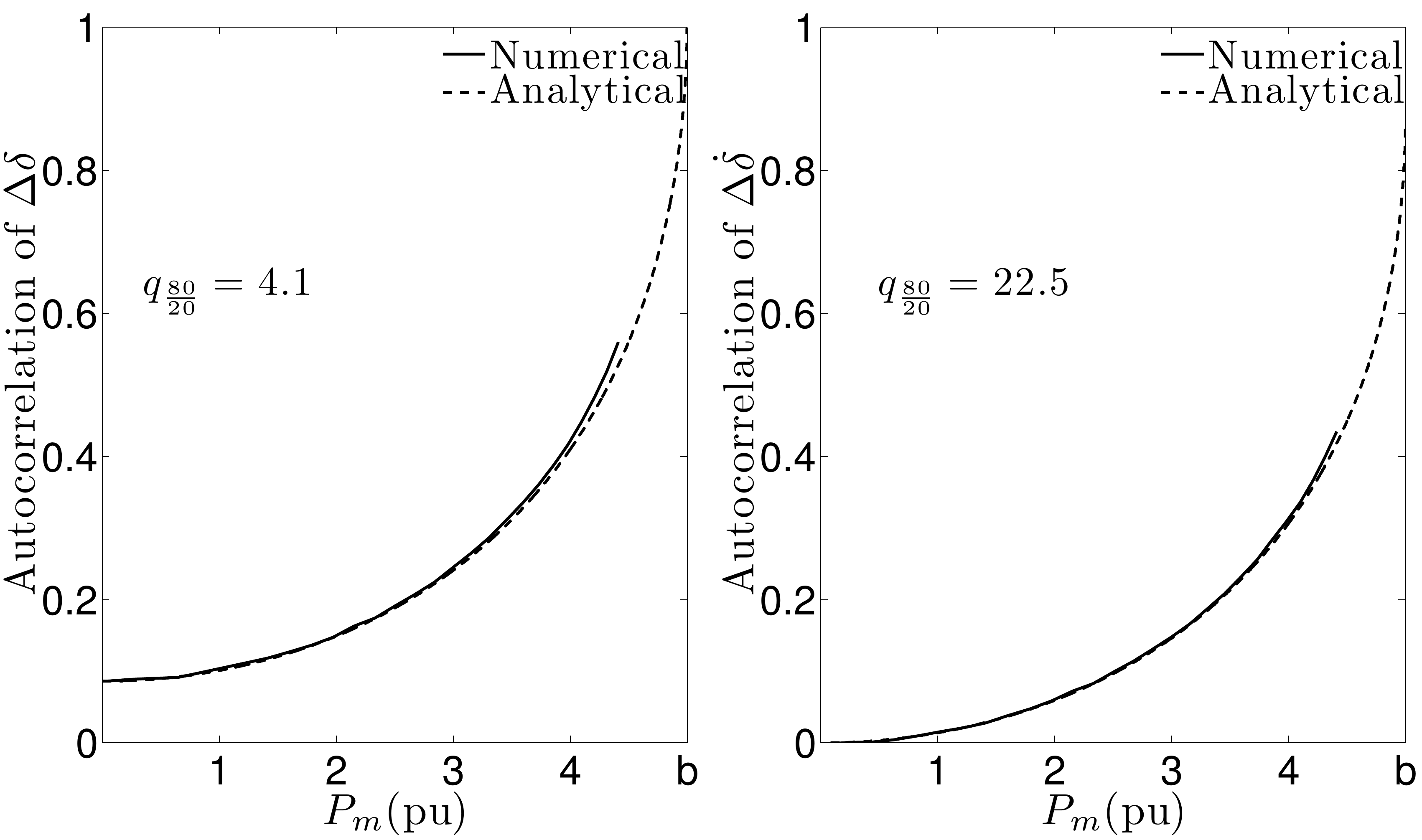}
\par\end{centering}
\vspace{-.15in}
\caption{\label{fig_sys1_ar_deom}Autocorrelations of $\Delta\delta,\Delta\dot{\delta}$
versus mechanical power $(P_{m})$ values. The autocorrelation values
are normalized by dividing by the variances of the variables. }

\vspace{-.1in}
\end{figure}

The results for the algebraic variables are mainly similar. Figs.~\ref{fig_sys1_var_vgtg},\ref{fig_sys1_ar_vgtg}
show the variances and autocorrelations of $\Delta V_{g},\Delta\theta_{g}$
as a function of load. In Fig.~\ref{fig_sys1_var_vgtg}, the variance
of $\Delta V_{g}$ decreases with $P_{m}$ until the system gets close
to the bifurcation, while the variance of $\Delta\theta_{g}$ increases
with $P_{m}$ even if the system is far from the bifurcation. The
autocorrelations of both $\Delta V_{g}$ and $\Delta\theta_{g}$ in
Fig.~\ref{fig_sys1_ar_vgtg} increase with $P_{m}$. However, the
ratio $q_{\frac{80}{20}}$ in~(\ref{eq:10-3}) is much larger for
$\Delta V_{g}$ than for $\Delta\theta_{g}$. This is caused by the
autocorrelation of $\Delta V_{g}$ being very close to zero for small
values of $P_{m}$. 

\begin{center}
\begin{figure}
\begin{centering}
\includegraphics[width=1\columnwidth,height=0.2\textheight]{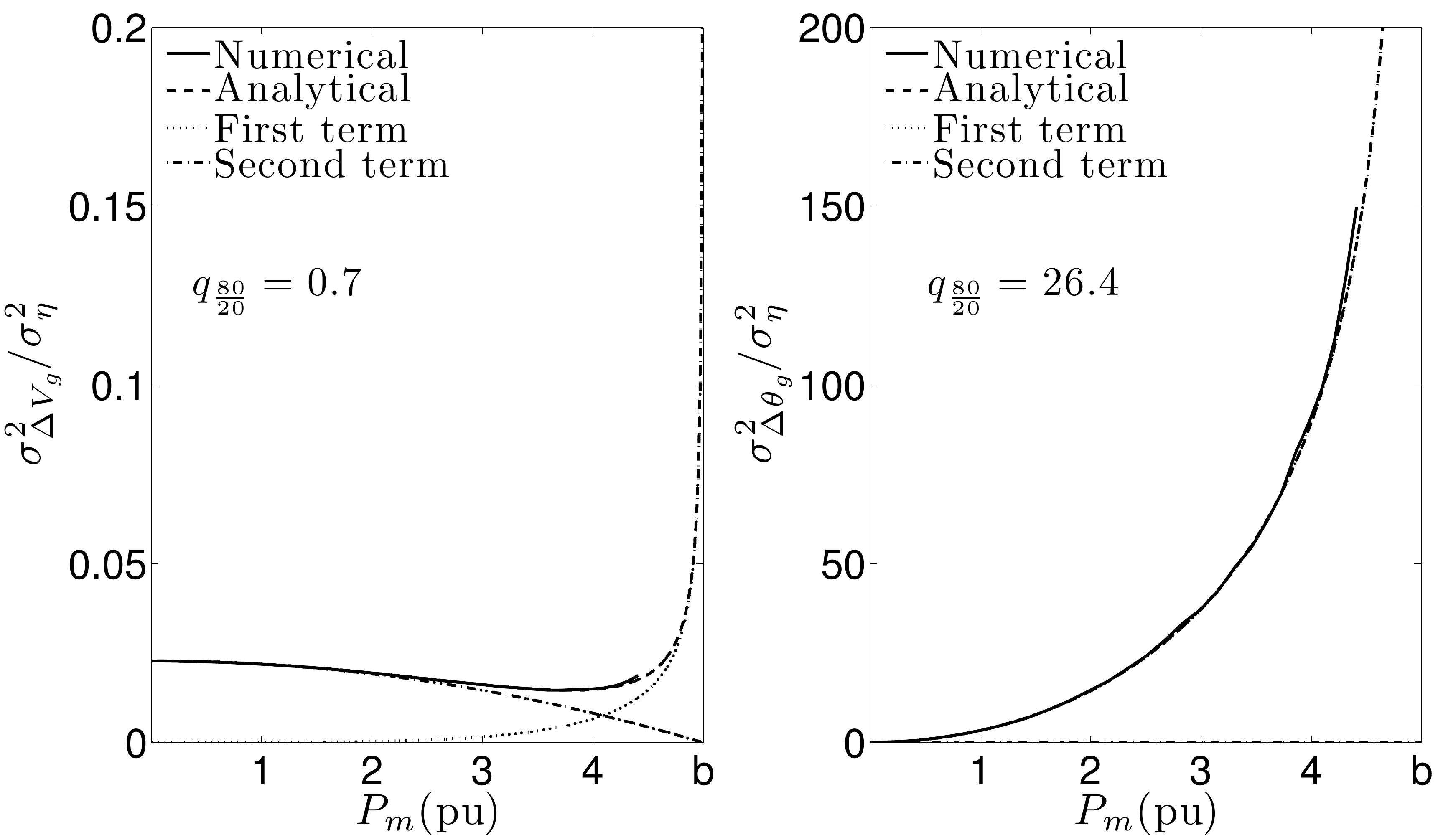}
\par\end{centering}
\vspace{-.15in}
\caption{\label{fig_sys1_var_vgtg}Variances of $\Delta V_{g}$ and $\Delta\theta_{g}$
versus mechanical power $\left(P_{m}\right)$ levels. The two terms
comprising the variances in (\ref{eq:24-3}) are also shown.}
\vspace{-.1in}
\end{figure}

\par\end{center}

\begin{center}
\begin{figure}
\begin{centering}
\includegraphics[width=1\columnwidth,height=0.2\textheight]{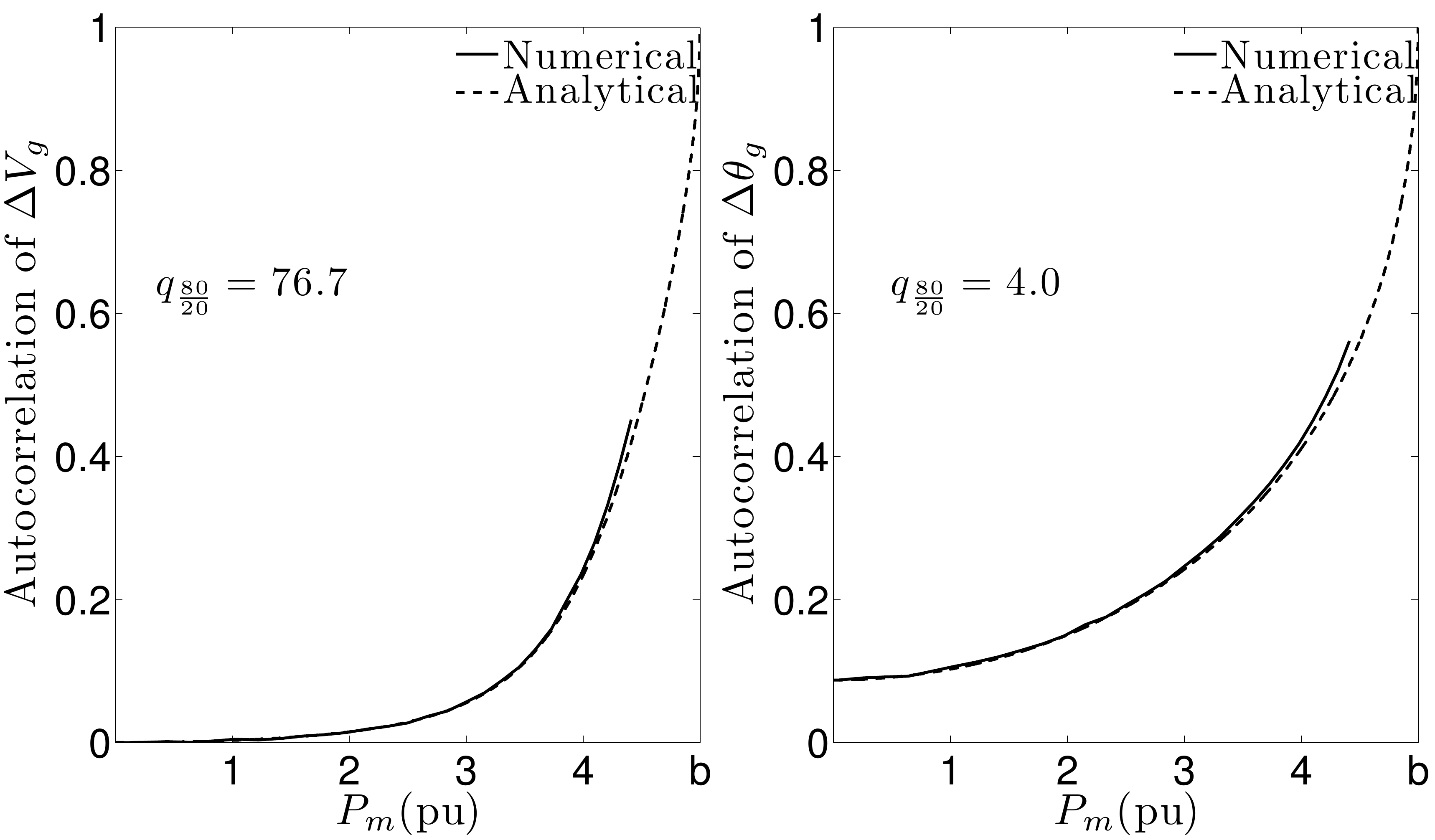}
\par\end{centering}
\vspace{-.15in}
\caption{\label{fig_sys1_ar_vgtg}Autocorrelations of $\Delta V_{g}$ and $\Delta\theta_{g}$
versus $P_{m}$.}

\vspace{-.2in}
\end{figure}

\par\end{center}

\vspace{-0.7in}

\subsection{Discussion\label{sub:SMIB_Discussion}}

These results can be better understood by observing the trajectory
of the eigenvalues of the SMIB system (Fig.~\ref{sys1_eigs}). Near
the bifurcation, the eigenvalues are very sensitive to changes in
the bifurcation parameter. As a result, the system is in the overdamped
regime $(\omega_{0}<\gamma)$ for much less than 0.1\% distance in
terms of $P_{m}$ to the critical transition. This implies that, at
least for this system, the autocorrelation function in \cite{Mitautocorrelation,podolsky2013critical},
is valid only when the system is within $0.1\%$ of the saddle-node
bifurcation. Because the method in \cite{Mitautocorrelation,podolsky2013critical}
can provide a good estimate of the autocorrelations and variances
of  state variables only for a very short range of the bifurcation
parameter, it may not be particularly useful as an early warning sign
of bifurcation. 

\begin{figure}
\vspace{-0.2in}
\begin{centering}
\includegraphics[width=1\columnwidth]{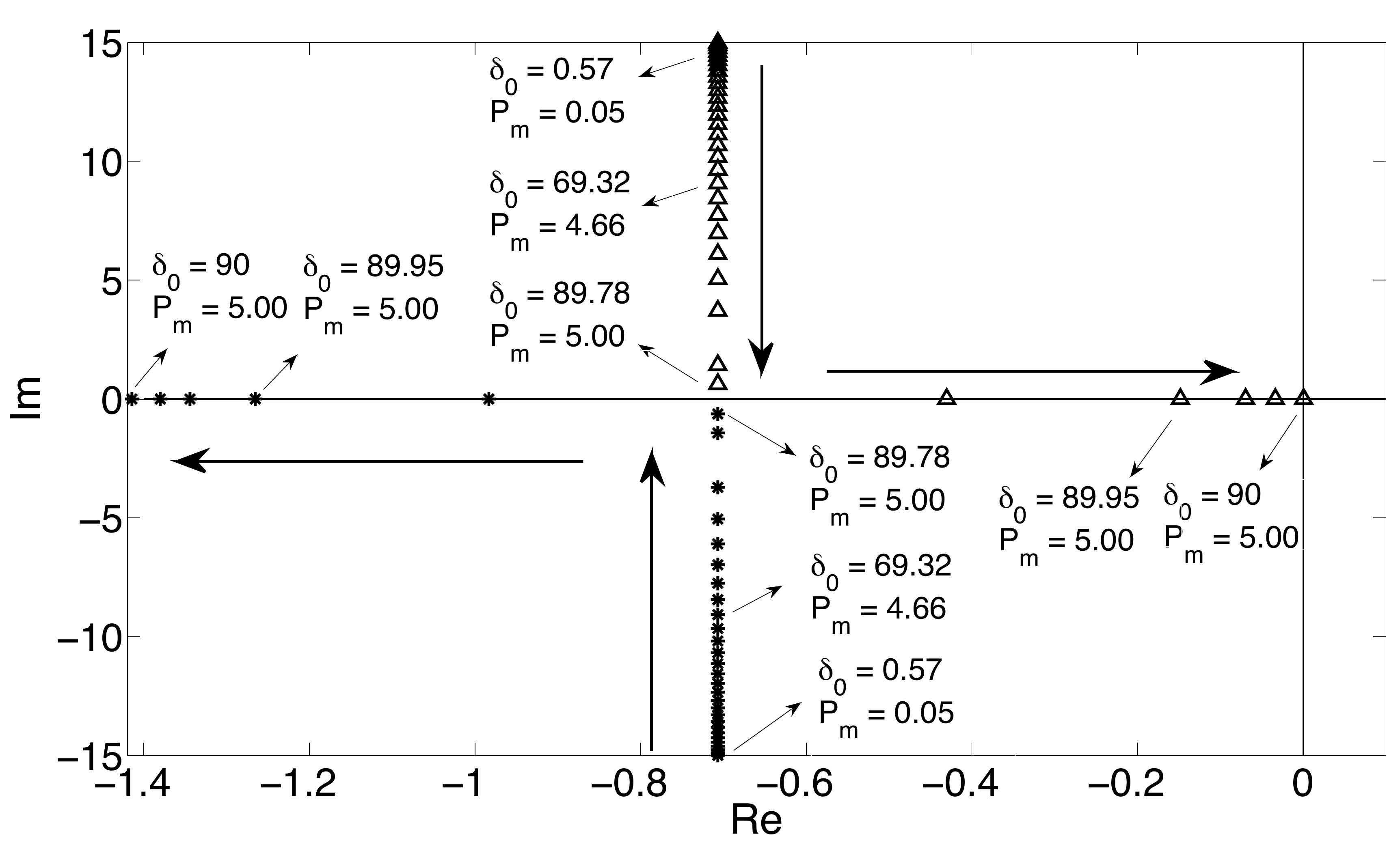}
\par\end{centering}
\vspace{-0.2in}
\caption{\label{sys1_eigs}Eigenvalues of the first system as the bifurcation
parameter (mechanical power) is increased. The arrows show the direction
of the eigenvalues' movement in the complex plane as $P_{m}$ is increased.
The values of $P_{m}$ and $\delta_{0}$ are given for several eigenvalues.}

\vspace{-.2in}
\end{figure}

From Figs.~\ref{fig_sys1_var_deom}--\ref{fig_sys1_ar_vgtg}, we
can observe that, except for the variance of $\Delta V_{g}$, the
variances and autocorrelations of all state variables increase when
the system is more loaded. This demonstrates that CSD occurs in this
system as it approaches bifurcation, as suggested both by general
results \cite{kuehn2011mathematical}, and prior work for power systems
\cite{CSDjournal,Mitautocorrelation}. 

In addition to validating these prior results, several new observations
can be made. For example, the signs of CSD are more clearly observable
in some variables than in others. While all of the variables show
some increase in autocorrelation and variance, they are less clearly
observable in $\Delta V_{g}$. The variance of $\Delta V_{g}$ decreases
with $P_{m}$ slightly until the vicinity of the bifurcation. In comparison,
the variance of $\Delta\theta_{g}$ always increases with $P_{m}$.
Fig.~\ref{fig_sys1_var_vgtg} shows the two terms of the expressions
for the variances of $\Delta V_{g}$ and $\Delta\theta_{g}$ in (\ref{eq:24-3}).
The second term of the variance of $\Delta\theta_{g}$ is very small
compared to the first term, and the first term is always dominant
and growing. On the other hand, the second term of the variance of
$\Delta V_{g}$ is more significant for small $P_{m}$. This term
decreases with $P_{m}$, which can be observed from the expression
for $C_{1_{,}2}$ in (\ref{eq:11-5}). Accordingly, decrease of $C_{1_{,}2}$
with $P_{m}$ causes the the variance of $\Delta V_{g}$ to decrease
with $P_{m}$ until the vicinity of the bifurcation. In conclusion,
the variance of $\Delta\theta_{g}$ is a better indicator of proximity
to the bifurcation. Because the variables $\Delta\delta$ and $\Delta\dot{\delta}$
are highly correlated with $\Delta\theta_{g}$, their variances are
also good indicators of proximity to the bifurcation.

The rate at which autocorrelation increases with $P_{m}$ differs
significantly in Figs.~\ref{fig_sys1_ar_deom} and \ref{fig_sys1_ar_vgtg}.
In Fig.~\ref{fig_sys1_ar_deom}, the ratio $q_{\frac{80}{20}}$ in
(\ref{eq:10-3}) is \textcolor{black}{5.5} times larger for $\Delta\dot{\delta}$
than for $\Delta\delta$. The normalized autocorrelation functions
of $\Delta\delta$ and $\Delta\dot{\delta}$ are as follows:
\begin{eqnarray}
\textnormal{{E}}\left[\Delta\delta\left(t\right)\Delta\delta\left(s\right)\right]/\sigma_{\Delta\delta}^{2} & = & \exp\left(-\gamma\Delta t\right)\frac{\omega_{0}}{\omega'}\label{eq:26-1}\\
 &  & \cdot\sin\left(\omega'\Delta t+\phi\right)\nonumber \\
\textnormal{{E}}\left[\Delta\dot{\delta}\left(t\right)\Delta\dot{\delta}\left(s\right)\right]/\sigma_{\Delta\dot{\delta}}^{2} & = & \exp\left(-\gamma\Delta t\right)\frac{\omega_{0}}{\omega'}\\
 &  & \cdot\sin\left(\omega'\Delta t+\pi-\phi\right)\nonumber 
\end{eqnarray}
The difference between the two functions is in the phase of the sine
function which causes the values of the two autocorrelations to be
different. $q_{\frac{80}{20}}$ is so much larger for $\Delta\dot{\delta}$
than for $\Delta\delta$ because of the time lag ($\Delta t$) used
to compute autocorrelation. $\Delta t=0.1\mbox{s}$ is close to the
zero crossing of the autocorrelation function of $\Delta\dot{\delta}$,
causing the large $q_{\frac{80}{20}}$.  This difference illustrates
the importance of choosing an appropriate time lag.

It is important to note that although the growth ratio of the autocorrelation
for $\Delta\delta$ is not large compared to $\Delta\dot{\delta}$,
it can be increased by subtracting a bias value from the autocorrelation
values for $P_{m}=0.2b(\textrm{{pu})}$ and $P_{m}=0.8b\textrm{({pu}})$.
For example, if the value of 0.075 is subtracted from the autocorrelation
values, the ratio $q_{\frac{80}{20}}$ increases from 4.1 to 13.0.
When using this approach, it is recommended that the new base value
(here, autocorrelation of $\Delta\delta$ for $P_{m}=0.2\textrm{b}(\textrm{{pu})}$)
is chosen to be at least $25\%$ of the original value, in order to
reduce the impact of measurement noise.

The results also show that it is the nonlinearity of this system that
causes CSD to occur. One of the elements of the state matrix $(-\omega_{0}^{2})$
in (\ref{eq:4-1}) changes with $P_{m}$ because of the nonlinear
relationship between the electrical power $(P_{g})$ and the rotor
angle, causing the eigenvalues to change with $P_{m}$. If the relationship
between $P_{g}$ and $\delta$ were linear, the state matrix $A$
would be constant.\textcolor{black}{{} }Indeed,\textcolor{black}{{} in
\cite{Gardiner:2004}, it is shown that the }stationary time correlation
matrix \textcolor{black}{of (\ref{eq:1-1})} can be calculated using
the following equation\textcolor{black}{:}
\begin{equation}
\textnormal{{E}}\left[\mathbf{\mathbf{\underline{\mathrm{\mathit{Z}}}\left(\mathit{t}\right)}}\underline{Z}{}^{T}\left(s\right)\right]=\exp\left[-A\Delta t\right]\sigma\label{eq:26-2}
\end{equation}
where $\sigma$ is the covariance matrix of the state variables.\textcolor{black}{{}
}Thus, t\textcolor{black}{he normalized autocorrelation matrix depends
only on $A$ and the time lag}. As a result, if the state matrix is
constant, the autocorrelation for a specific time lag will also be
constant. Thus, in this system, CSD is caused by the nonlinear relationship
between $P_{g}$ and the rotor angle.

\vspace{-.15in}
\section{Single Machine Single Load System\label{sec:voltage-collapse}}

The first system illustrates how CSD can occur in a generator connected
to a large power grid, through a long line. In this section we use
a generator to represent the bulk grid, and look for signs of CSD
caused by a stochastically varying load. Some form of the single machine
single load (SMSL) model used in this section has been used extensively
to study voltage collapse (e.g., \cite{canizares2002voltage,kundur1994power}).

\vspace{-.15in}
\subsection{Stochastic SMSL System Model \label{sub1:voltage_collapse}}

The second system (shown in Fig.~\ref{SMSL}) consists of one generator,
one load and a transmission line between them. The random variable
$\eta$ defined in (\ref{eq:2-3}) and (\ref{eq:2-4}), is added to
the load to model its fluctuations. The load consists of both active
and reactive components. In order to stress the system, the baseline
load $S_{d}$ is increased, while keeping the noise intensity ($S_{d0}$)
and the load's power factor constant.
\vspace{-.15in}
\begin{center}
\begin{figure}
\begin{centering}
\includegraphics[width=1\columnwidth]{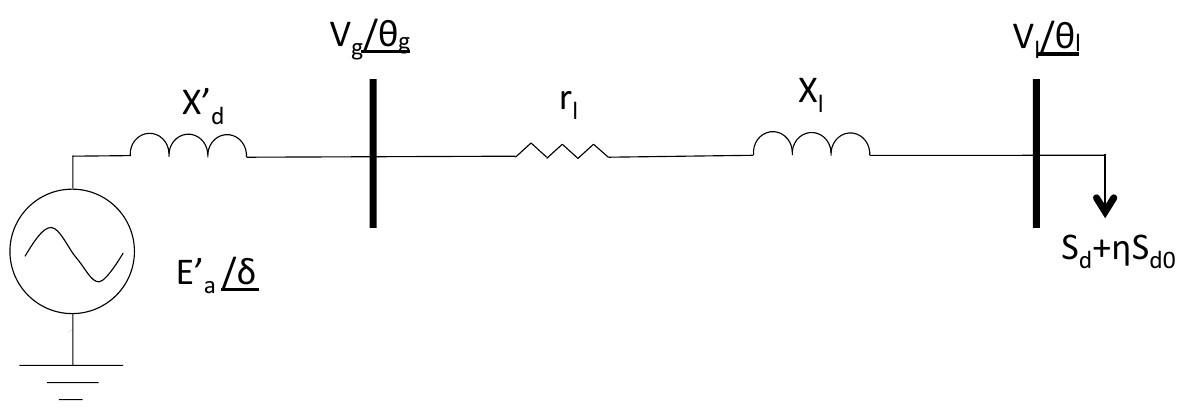}
\par\end{centering}

\caption{\label{SMSL}Single machine single load system. }

\vspace{-.1in}
\end{figure}

\par\end{center}
\vspace{-.1in}
A set of differential-algebraic equations comprising the swing equation
and power flow equations describe this system. The swing equation
and the generator's electrical power equation are given below:
\begin{eqnarray}
M\ddot{\delta}+D\dot{\delta} & = & P_{m}-P_{g}\label{eq:26}\\
P_{g} & = & E_{a}^{'}V_{l}G_{gl}\cos\left(\delta-\theta_{l}\right)\label{eq:27}\\
 &  & +E_{a}^{'}V_{l}B_{gl}\sin\left(\delta-\theta_{l}\right)+E_{a}^{'2}G_{gg}\nonumber 
\end{eqnarray}
where $V_{l},\theta_{l}$ are voltage magnitude and angle of the load
busbar, $G_{gl},G_{gg}$ and $B_{gl}$ are as follows:
\begin{eqnarray}
G_{gg} & = & -G_{gl}=\textrm{\ensuremath{\mathsf{{Re}}}}\left(\frac{1}{r_{l}+jX_{l}}\right)\label{eq:28}\\
B_{gl} & = & \textrm{\ensuremath{\mathsf{-{Im}}\left(\frac{1}{r_{l}+jX_{l}}\right)}}\label{eq:29}
\end{eqnarray}

The power flow equations at the load bus are as follows:
\begin{eqnarray}
-P_{d}-P_{d0}\eta & = & V_{l}E_{a}^{'}G_{gl}\cos\left(\theta_{l}-\delta\right)\label{eq:31}\\
 &  & +V_{l}E_{a}^{'}B_{gl}\sin\left(\theta_{l}-\delta\right)+V_{l}^{2}G_{ll}\nonumber 
\end{eqnarray}
\begin{eqnarray}
-Q_{d}-Q_{d0}\eta & = & V_{l}E_{a}^{'}G_{gl}\sin\left(\theta_{l}-\delta\right)\label{eq:32}\\
 &  & -V_{l}E_{a}^{'}B_{gl}\cos\left(\theta_{l}-\delta\right)-V_{l}^{2}B{}_{ll}\nonumber 
\end{eqnarray}
where $G_{ll}=G_{gg},\, B_{ll}=-B_{gl}$, and $P_{d0},Q_{d0}$ are
constant values. The parameters of this system are similar to the
SMIB system, with the following additional parameters: $r_{l}=0.025\Omega,P_{d0}=1\textrm{pu}$,
$pf=0.95\textrm{{lead}}$, where $r_{l}$ is the line's resistance
and $pf$ is the load's power factor.

In this system, $V_{l}$, $\theta_{l}-\delta$ are the algebraic variables,
and $\delta$, $\dot{\delta}$ are the differential variables. The
algebraic equations (\ref{eq:31}) and (\ref{eq:32}) define $V_{l}$
and $\theta_{l}-\delta$, which then drive $\delta$ through (\ref{eq:26})
and (\ref{eq:27}). By linearizing (\ref{eq:27}) and the power flow
equations around the equilibrium, we simplified (\ref{eq:26}) to
the following:
\begin{equation}
\Delta\ddot{\delta}+\frac{D}{M}\Delta\dot{\delta}=-\frac{C_{5}}{M}\eta\label{eq:35-7}
\end{equation}
where $C_{5}$ is a function of the system state at the equilibrium
point. The derivation of (\ref{eq:35-7}) and the expression for $C_{5}$
are presented in Appendix \ref{appendix:SMSL}. Comparing (\ref{eq:2-5})
with (\ref{eq:35-7}) yields the following:
\begin{equation}
\gamma=\frac{D}{2M},\omega_{0}=0,f=\frac{C_{5}}{M}\label{eq:4-8}
\end{equation}

The expression for the autocorrelation of $\Delta\dot{\delta}$ is
given in (\ref{eq:38-2}). Note that the normalized autocorrelation
of $\Delta\dot{\delta}$ does not change with the bifurcation parameter
$(P_{d})$, as it did for the SMIB system. In Appendix \ref{appendix:SMSL},
it is shown that $\Delta V_{l}$ and $\Delta\delta-\Delta\theta_{l}$
are proportional to $ $$\eta$ (see (\ref{eq:33}) and (\ref{eq:34})).
As a result, they are memoryless; the variables have zero autocorrelation. 

Figs.~\ref{fig_sys2_var_om} and \ref{fig_sys2_var_devl} show the
analytical and numerical solutions of the variances of $\Delta\dot{\delta}$,
$\Delta V_{l}$ and $\Delta\delta-\Delta\theta_{l}$. Unlike the SMIB
system, the variance of $\Delta V_{l}$ is a good early warning sign
of the bifurcation. It is also much more sensitive to the increase
of $P_{d}$ compared to $\Delta\delta-\Delta\theta_{l}$ and $\Delta\dot{\delta}$. 

\begin{center}
\begin{figure}
\begin{centering}
\includegraphics[width=1\columnwidth,height=0.2\textheight]{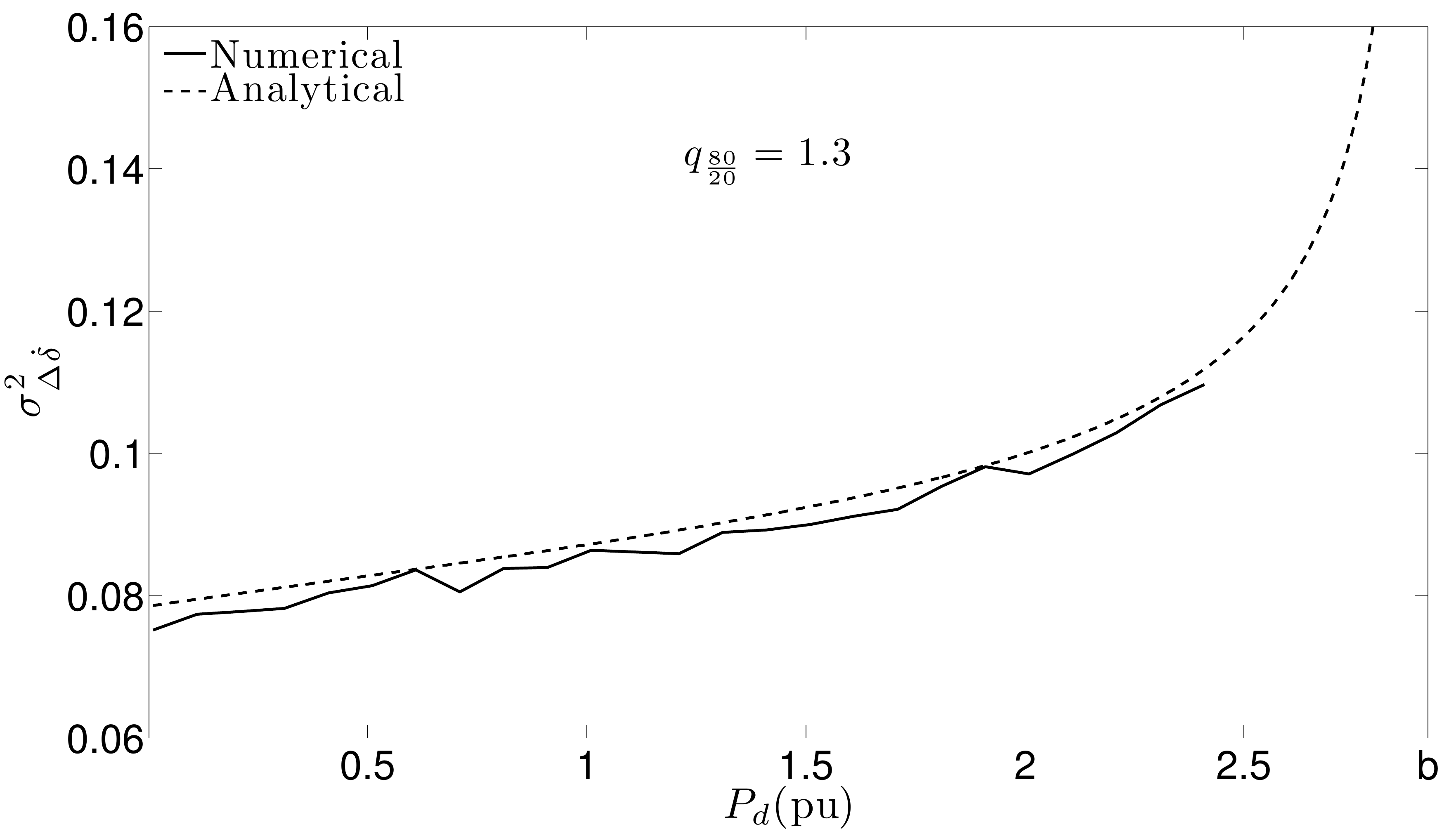}
\par\end{centering}

\caption{\label{fig_sys2_var_om}Variance of $\Delta\dot{\delta}$ for different
load levels. The variance increases modestly with $P_{d}$ as the
system approaches the bifurcation.}

\vspace{-.1in}
\end{figure}

\par\end{center}

\begin{center}
\begin{figure}
\begin{centering}
\includegraphics[width=1\columnwidth,height=0.2\textheight]{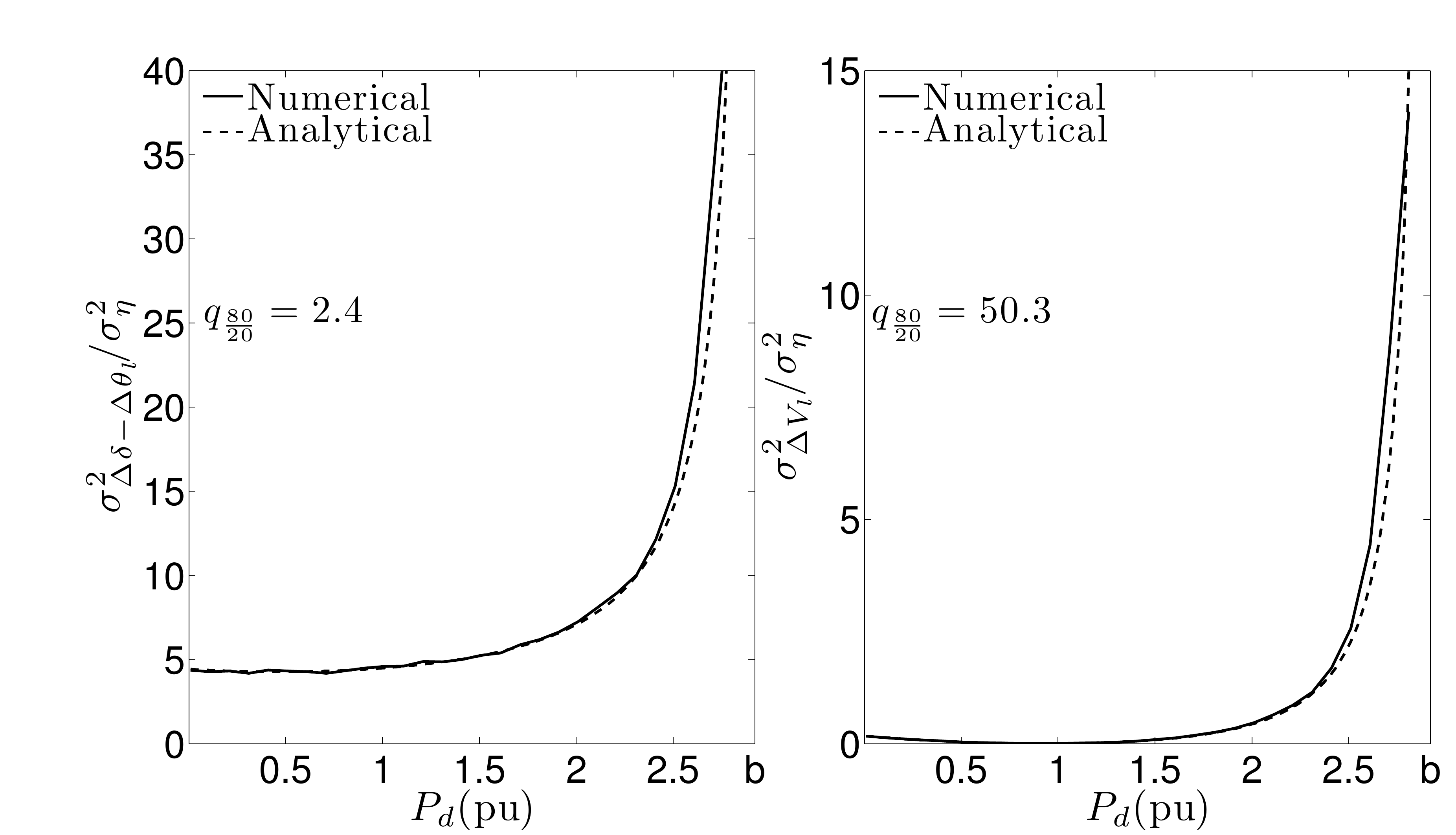}
\par\end{centering}

\caption{\label{fig_sys2_var_devl}Variances of $\Delta\delta-\Delta\theta_{l}$
and $\Delta V_{l}$ for different load levels. Both variances increase
with $P_{d}$ as the system approaches the bifurcation.}

\vspace{-.25in}
\end{figure}

\par\end{center}

\vspace{-0.7in}
\subsection{Discussion}

As was the case with the SMIB system, when the power flowing on the
transmission line in this system reaches its transfer limit, the algebraic
equations become singular. However, unlike the previous system, the
differential equations of this system do not become singular at the
bifurcation point of the algebraic equations. Fig.~\ref{fig_sys2_traj}
shows the sample trajectories of the two systems' rotor angles. Both
signals are Gaussian stochastic processes. The rotor angle in the
SMIB system is an Ornstein--Uhlenbeck process while the rotor angle
in the SMSL system varies like the position of the brownian particle
\cite{Noise-induced}. The existence of the infinite bus in the former
system causes this difference. 
\vspace{-0.2in}
\begin{center}
\begin{figure}
\vspace{-.15in}
\begin{centering}
\includegraphics[width=1\columnwidth,height=0.2\textheight]{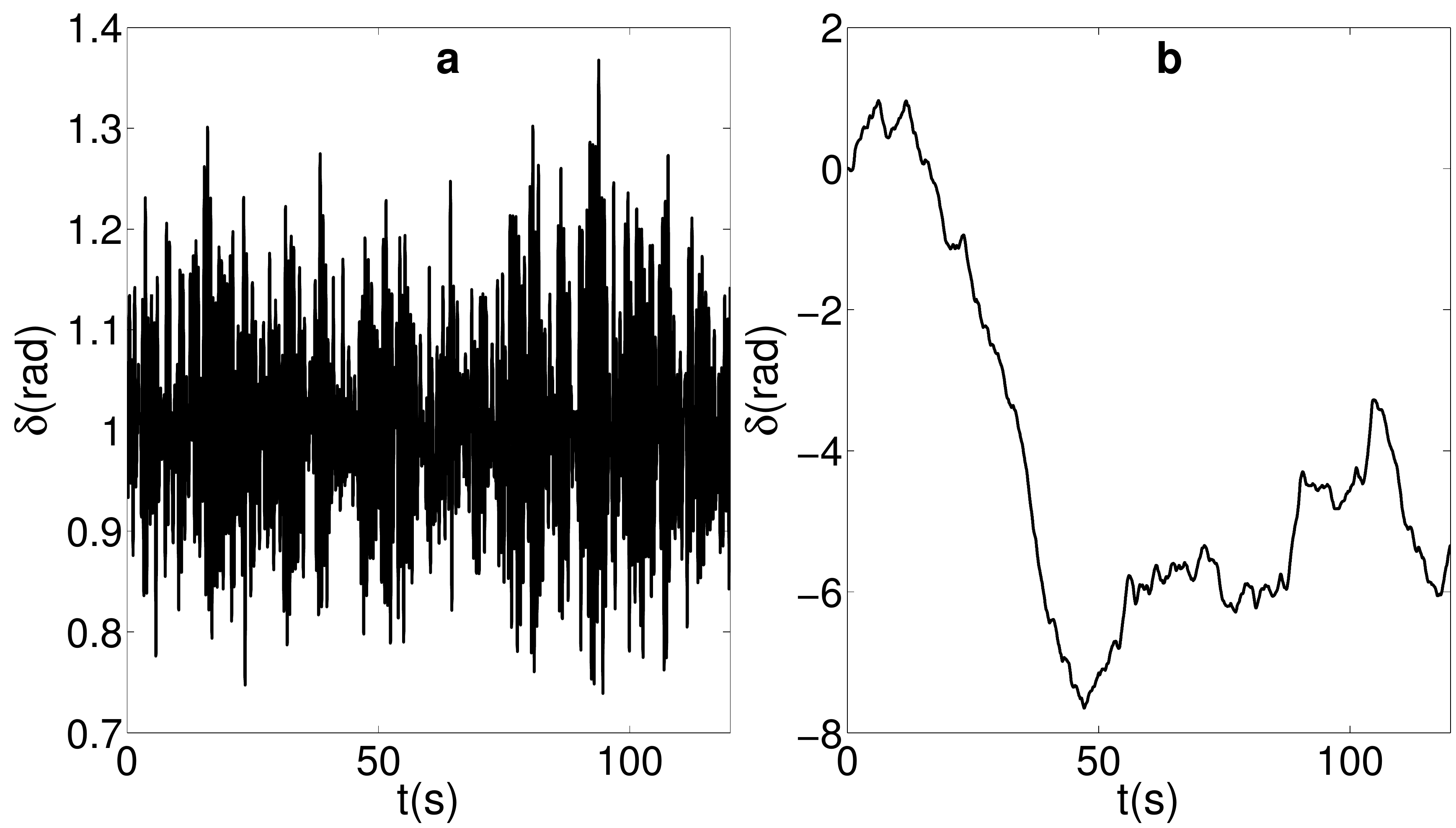}
\par\end{centering}

\caption{\label{fig_sys2_traj}A sample trajectory of the rotor angle of (a)
the SMIB system (b) the SMSL system. }
\end{figure}

\par\end{center}

One difference between the SMSL system and the SMIB system is the
absence of the term comprising $\Delta\delta$ in (\ref{eq:35-7})
compared with (\ref{eq:2-5}). This causes the linearized state matrix
to be independent of the bifurcation parameter\textcolor{black}{.
From (\ref{eq:26-2}), one can show that the normalized autocorrelation
of $\Delta\dot{\delta}$  depends only on $A$ and the time lag. Since
$A$} is constant in this system, \textcolor{black}{the autocorrelation
of $\Delta\dot{\delta}$} will be constant for a specific $\Delta t$. 

The increase of the variances of both differential and algebraic variables
is due to the non-linearity of the algebraic equations. Fig. \ref{dvdp_fig}
shows that as the load power increases, the perturbation of the load
power causes a larger deviation in the load busbar voltage magnitude.
Consequently, variance of this algebraic variable increases with $P_{d}$.
Likewise, this nonlinearity causes the coefficient $C_{5}$ in (\ref{eq:35-7})
to increase as the load power is increased, increasing the variance
of $\Delta\dot{\delta}$. One can show that if the line resistance
($r_{l}$) is neglected in this system, $C_{5}$ in (\ref{eq:35-7})
will be replaced by $P_{d0}$. In this case, the variance of $\Delta\dot{\delta}$
is constant, since the differential and algebraic equations are fully
decoupled.

\begin{figure}
\includegraphics[width=1\columnwidth]{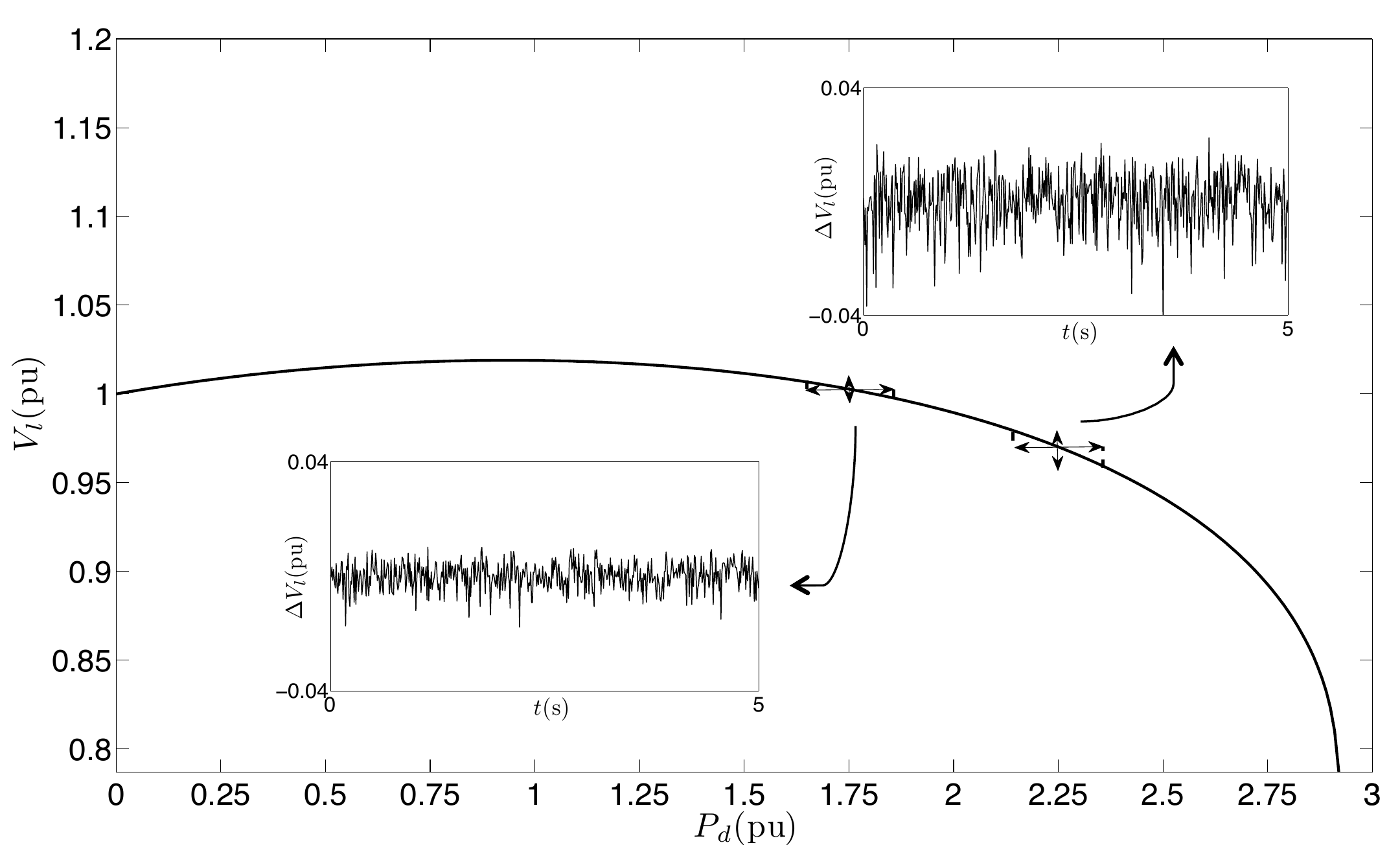}

\caption{\label{dvdp_fig}The load bus voltage as a function of load power.
The load bus voltage magnitude becomes increasingly sensitive to power
fluctuations as the system approaches the bifurcation. This increased
sensitivity raises the voltage magnitude's variance.}

\vspace{-.25in}
\end{figure}

\textcolor{black}{While voltage variance increases with load, this
system does not technically show CSD before the bifurcation, since
increases in both variance and autocorrelation are essential to conclude
that CSD has occurred \cite{kuehn2011mathematical}. Also, the eigenvalues
of the state matrix of this system do not vary with load. This confirms
that CSD does not occur in this system, since the poles of the dynamical
system do not move toward the right-half plane as the bifurcation
parameter increases \cite{scheffer2009early}, \cite{kuehn2011mathematical}. }

\vspace{-0.1in}

\section{Three-Bus System\label{sec:Three-Bus-System}}
\vspace{-0in}
Real power systems have properties that are common to both the SMIB
in Sec.~\ref{sec:SMIB} and the SMSL in Sec.~\ref{sec:voltage-collapse}.
In order to explore CSD for a system that has both an infinite bus,
and the potential for voltage collapse case, this section looks at
the three-bus system in Fig.~\ref{Three bus}. 

\vspace{-0.15in}
\subsection{Model and Results}

The three-bus system  consists of a generator connected to a load
bus through a transmission line, which is  connected to an infinite
bus through another transmission line. In the SMIB system, the bifurcation
occurred in the differential equations. Increasing the load in the
three-bus system causes a saddle-node bifurcation in the algebraic
equations $F_{1}\left(\delta,\underline{y},0\right)=0,\underline{F_{2}}\left(\delta,\underline{y},0\right)=0$
(in terms of (\ref{eq:2-1}), (\ref{eq:2-2})), as in the SMSL system.
However, unlike in the SMSL system, the bifurcation in these algebraic
equations also causes a bifurcation in the differential equation (\ref{eq:2-5}). 
\vspace{-.15in}
\begin{center}
\begin{figure}[H]
\begin{centering}
\includegraphics[width=1\columnwidth]{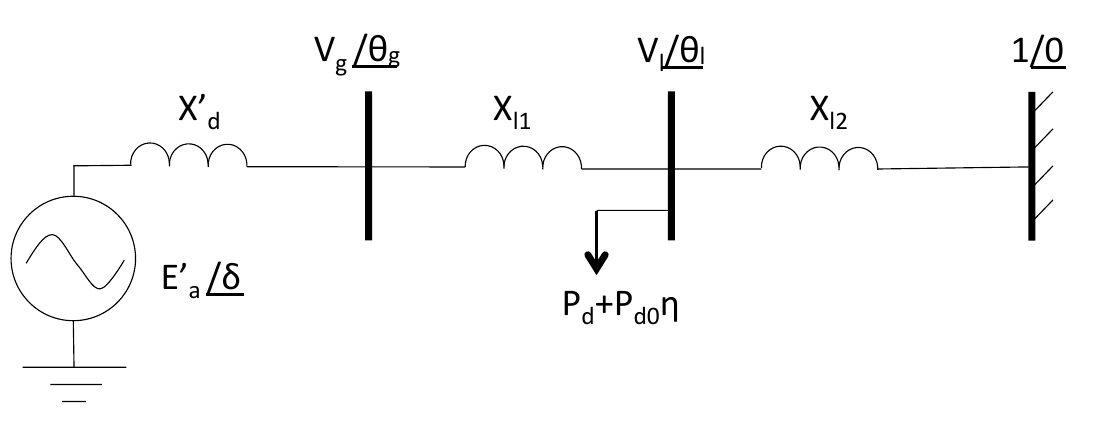}
\par\end{centering}

\caption{\label{Three bus}Three--bus system. }

\vspace{-.3in}
\end{figure}

\par\end{center}

We studied this system for two different cases. Our goal from studying
these two cases was to show that the CSD signs for some variables
can vary differently with changing the system parameters. In case
A, the parameters of this system are similar to those in the SMIB
system except for the following: 

\begin{center}
$X_{l1}=\textrm{0.1\textrm{pu}},X_{l2}=0.35\textrm{pu},X'_{d}=0.1\textrm{pu}$ 
\par\end{center}

\begin{flushleft}
In Case~B, the following parameters were used: 
\par\end{flushleft}

\begin{center}
$X_{l1}=\textrm{0.3\textrm{pu}},\, D=0.001\frac{\textnormal{{pu}}}{rad/s}$
\par\end{center}

The algebraic equations of the three-bus system are as follows:
\begin{eqnarray}
\left(\frac{E'_{a}V_{l}}{X}\sin\left(\delta-\theta_{l}\right)-\frac{2}{3}P_{d}\right)/M & = & 0\label{eq:39}\\
\frac{E'_{a}V_{l}}{X}\sin\left(\delta-\theta_{l}\right)-\frac{V_{l}}{X_{l2}}\sin\left(\theta_{l}\right)-P_{d0}\eta & = & P_{d}\label{eq:40}\\
\frac{E'_{a}V_{l}}{X}\cos\left(\delta-\theta_{l}\right)+\frac{V_{l}}{X_{l2}}\cos\left(\theta_{l}\right) & = & V_{l}^{2}\cdot\label{eq:41}\\
 &  & \left(\frac{1}{X}+\frac{1}{X_{l2}}\right)\nonumber 
\end{eqnarray}
where $X=X_{d}^{'}+X_{l1}$, $V_{l},\theta_{l}$ are voltage magnitude
and angle of the load busbar. Equation (\ref{eq:39}) is equivalent
to $F_{1}\left(\delta,\underline{y},0\right)$ in (\ref{eq:2-1}),
and (\ref{eq:40}), (\ref{eq:41}), which are the simplified active
and reactive power flow equations at the load busbar, are equivalent
to $F_{2}\left(\delta,\underline{y},0\right)$ in (\ref{eq:2-2}).
We assumed that $P_{g0}=\nicefrac{2P_{d}}{3}$, which is reflected
in (\ref{eq:39}).

The following equalities relate this system to the general model in
(\ref{eq:2-5}):
\begin{equation}
\gamma=\frac{D}{2M};\omega_{0}^{2}=\frac{-C_{6}}{M};f=\frac{-C_{7}}{M}\label{eq:5-4}
\end{equation}
where $C_{6}$ and $C_{7}$ are functions of the system state at the
equilibrium point. The derivation and expressions for $C_{6},C_{7}$
are presented in Appendix \ref{appendix:sys3_1}. Fig. \ref{C6C7}
shows $C_{6},C_{7}$ versus $P_{d}$. When the load increases, $C_{6}$
approaches $0,$ and a bifurcation in the differential equation (\ref{eq:2-5})
and (\ref{eq:5-4}) occurs. 

\begin{figure}
\begin{centering}
\includegraphics[width=1\columnwidth]{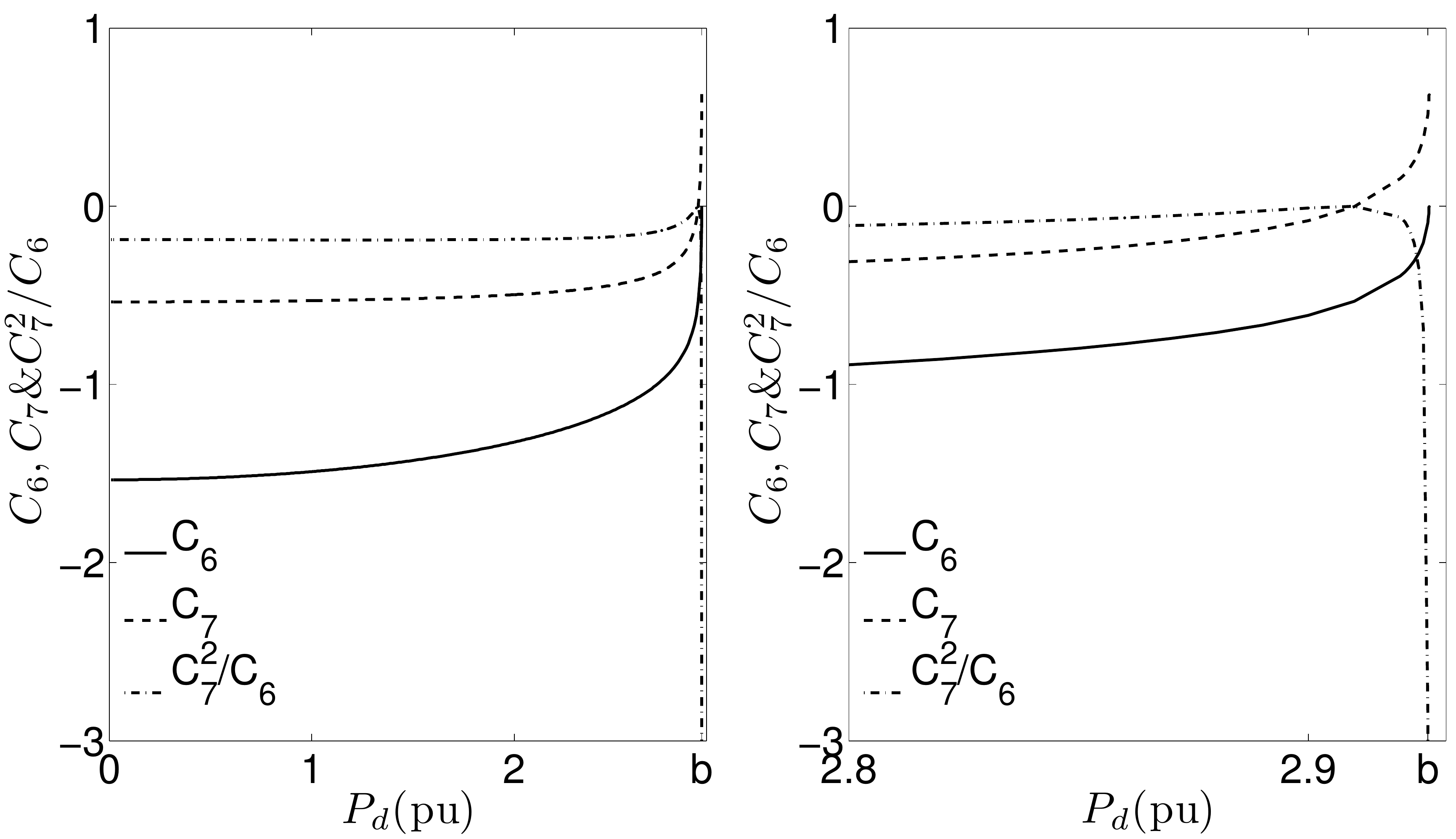}
\par\end{centering}
\vspace{-.15in}
\caption{\label{C6C7}Three variables $C_{6},C_{7}$ and $C_{7}^{2}/C_{6}$
derived by linearizing the Three-bus system model. The left panel
shows the variables versus $P_{d}$ for Case B. The right panel shows
a close-up view of the variables near the bifurcation. Note that as
$P_{d}\rightarrow P_{d,cr}$, $C_{6}\rightarrow0$ while $C_{7}$
approaches a finite value of $\sim0.6$. $C_{7}^{2}/C_{6}\rightarrow\infty$,
as $P_{d}\rightarrow P_{d,cr}$. }
\vspace{-.3in}
\end{figure}

Using~(\ref{eq:5-4}), the expressions in Sec.~\ref{sub:Calculation-of-Autocorrelations},
and (\ref{eq:52}), (\ref{eq:53}) in Appendix \ref{appendix:sys3_1},
we calculated the variances and autocorrelations of $\Delta\delta,\Delta\dot{\delta},\Delta V_{l}$
and $\Delta\theta_{l}$. We chose the autocorrelation time lag $\Delta t$
of the variables to be equal to 0.14s taking a similar approach as
in Sec.~\ref{sec: SMIB-2}. Although the chosen $\Delta t$ may not
be optimal for all of the variables, it represents a reasonable compromise
between simplicity (choosing just one $\Delta t$) and usefulness
as early warning signs. Figs. \ref{fig_sys3_var_deom}--\ref{fig_sys3_ar_vltl}
compare the analytical solutions with the numerical solutions of the
variances and autocorrelations of $\Delta\delta$, $\Delta\dot{\delta}$,
$\Delta V_{l}$ and $\Delta\theta_{l}$. 

Fig.~\ref{fig_sys3_ar_deom} shows that although the growth rates
of the autocorrelations of $\Delta\delta,\Delta\dot{\delta}$ are
not large, the autocorrelations increase monotonically in both cases.
As mentioned in Sec.~\ref{sub:SMIB_Discussion}, it is possible to
have larger indicators (growth ratios) by subtracting a bias value
from the autocorrelations. On the other hand, the variances of $\Delta\delta,\Delta\dot{\delta}$
in Fig. \ref{fig_sys3_var_deom}, do not monotonically increase for
case B. We will explain this behavior in the next subsection. As a
result, they are not reliable indicators of proximity to the bifurcation.

Fig.~\ref{fig_sys3_var_vltl} shows that although both variances
of $\Delta V_{l}$ and $\Delta\theta_{l}$ increase with $P_{d}$,
increase of the variance of $\Delta V_{l}$ is more significant. Also,
the variance of $\Delta\theta_{l}$ does not increase monotonically
with $P_{d}$ for case B. As a result, the variance of $\Delta V_{l}$
seems to be a better indicator of the system stability. 

In Fig.~\ref{fig_sys3_ar_vltl}, the autocorrelation of $\Delta V_{l}$
until very near the bifurcation is small compared to those in Fig.~\ref{fig_sys3_ar_deom}.
This is caused by $C_{26}$ being very small in (\ref{eq:52}), so
$\Delta V_{l}$ is tied to the differential variables weakly. As a
result, $\Delta V_{l}$ behaves in part like $\eta$---the white
random variable, and hence its autocorrelation is not a good indicator
of proximity to the bifurcation. In addition, nonmonotonicity of the
autocorrelations of $\Delta V_{l},\,\Delta\theta_{l}$ for case B
in Fig. \ref{fig_sys3_ar_vltl} shows that they are not good early
warning signs of bifurcation.

\begin{center}
\begin{figure}
\begin{centering}
\includegraphics[width=1\columnwidth,height=0.2\textheight]{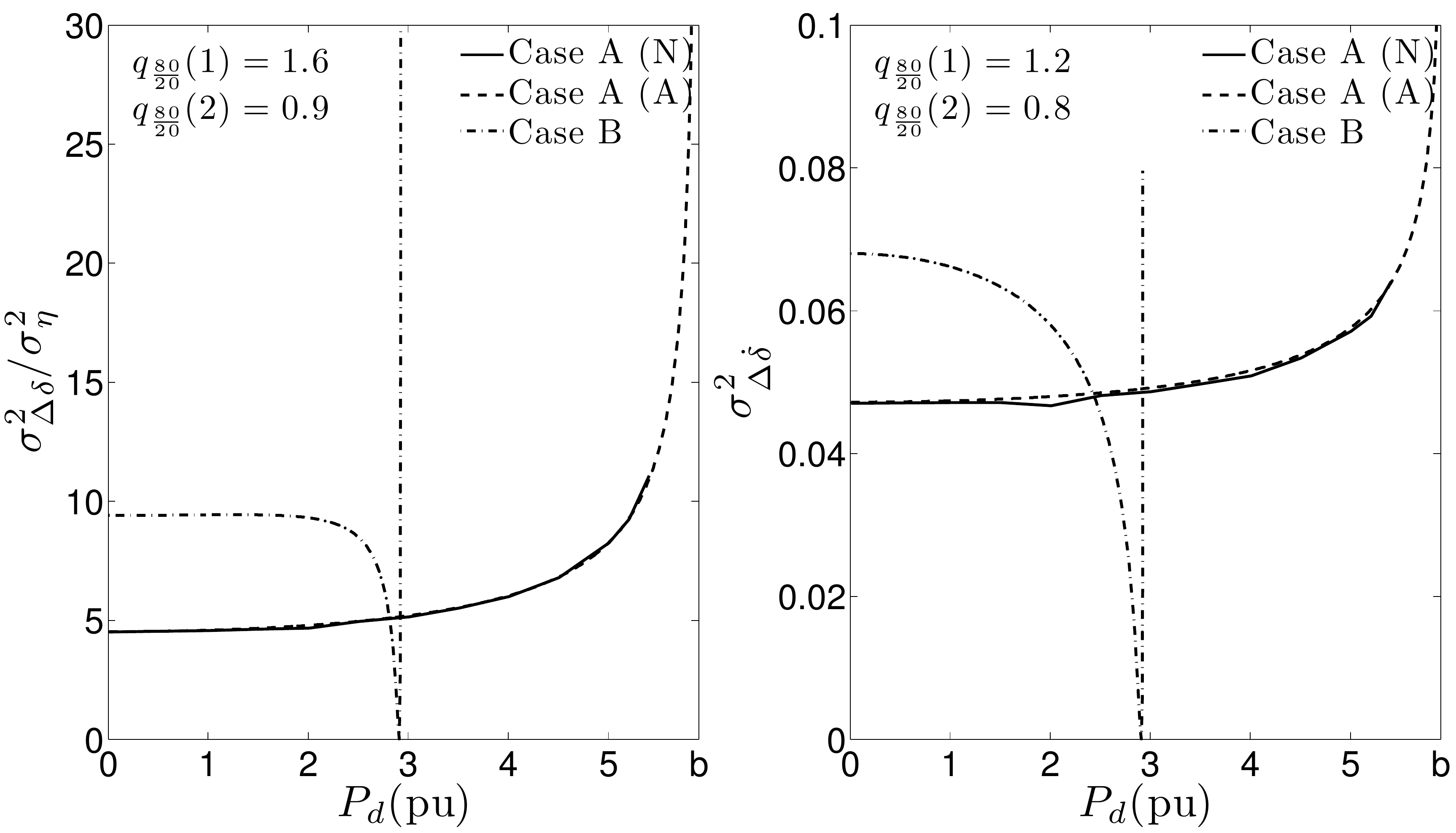}
\par\end{centering}

\caption{\label{fig_sys3_var_deom}Variances of $\Delta\delta,\Delta\dot{\delta}$
versus load power $\left(P_{d}\right)$. The ratios $q_{\frac{80}{20}}(1),q_{\frac{80}{20}}(2)$
are for case A, case B respectively. $ $CaseA(N), CaseA(A) denote
numerical and analytical solutions for case A.}

\vspace{-.2in}
\end{figure}

\par\end{center}

\begin{center}
\begin{figure}
\begin{centering}
\includegraphics[width=1\columnwidth,height=0.22\textheight]{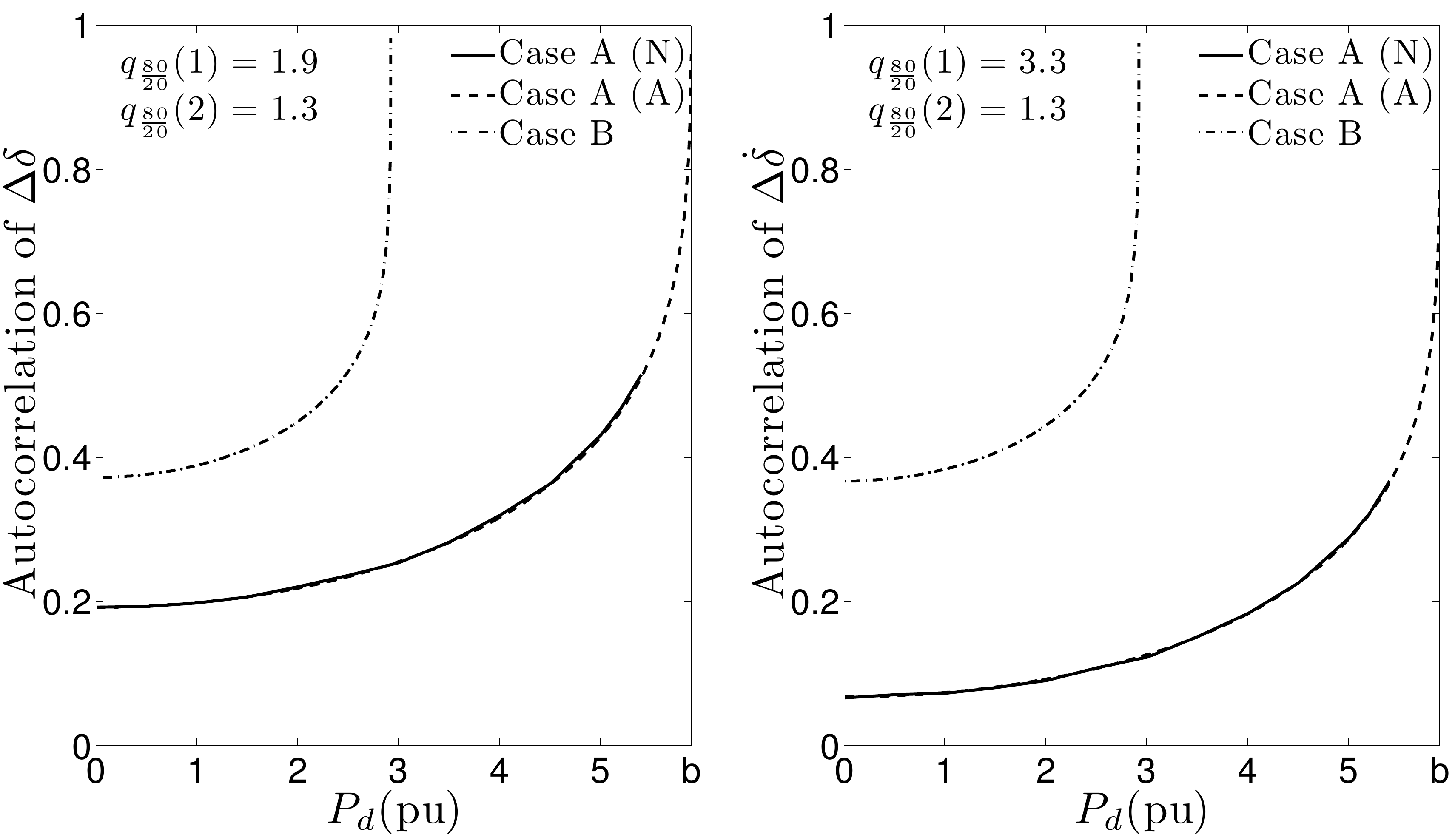}
\par\end{centering}
\vspace{-.1in}
\caption{\label{fig_sys3_ar_deom}Autocorrelations of $\Delta\delta,\Delta\dot{\delta}$
versus $\textrm{P}_{d}$. Both of the autocorrelations increase with
$P_{d}$. }

\vspace{-.1in}
\end{figure}
\begin{figure}
\begin{centering}
\includegraphics[width=1\columnwidth,height=0.22\textheight]{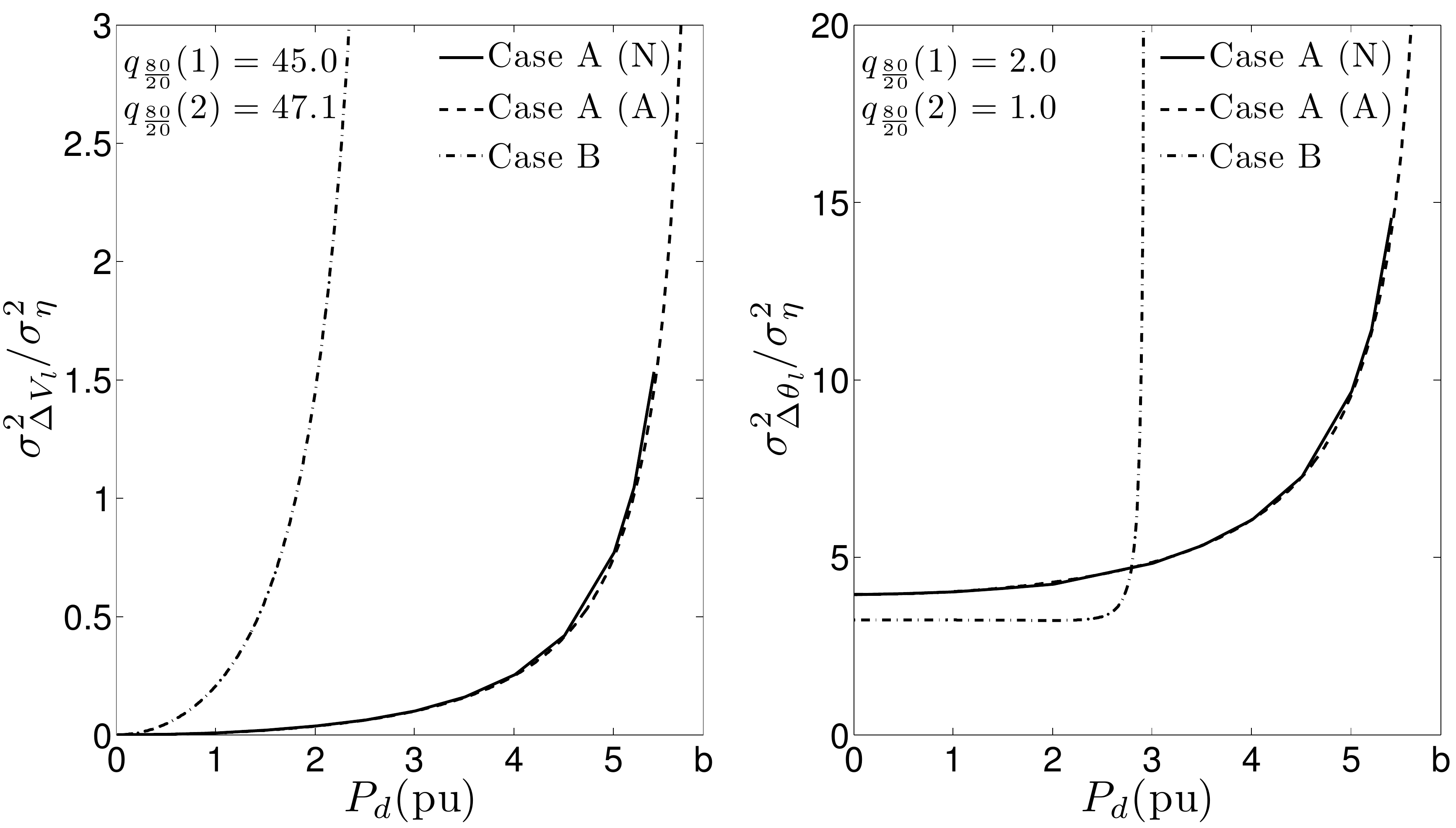}
\par\end{centering}
\vspace{-.1in}
\caption{\label{fig_sys3_var_vltl}Variances of $\Delta V_{l},\Delta\theta_{l}$
versus $P_{d}$. }

\vspace{-.0in}
\end{figure}

\par\end{center}
\vspace{-0.1in}
\begin{center}
\begin{figure}
\vspace{-0.15in}
\begin{centering}
\includegraphics[width=1\columnwidth,height=0.22\textheight]{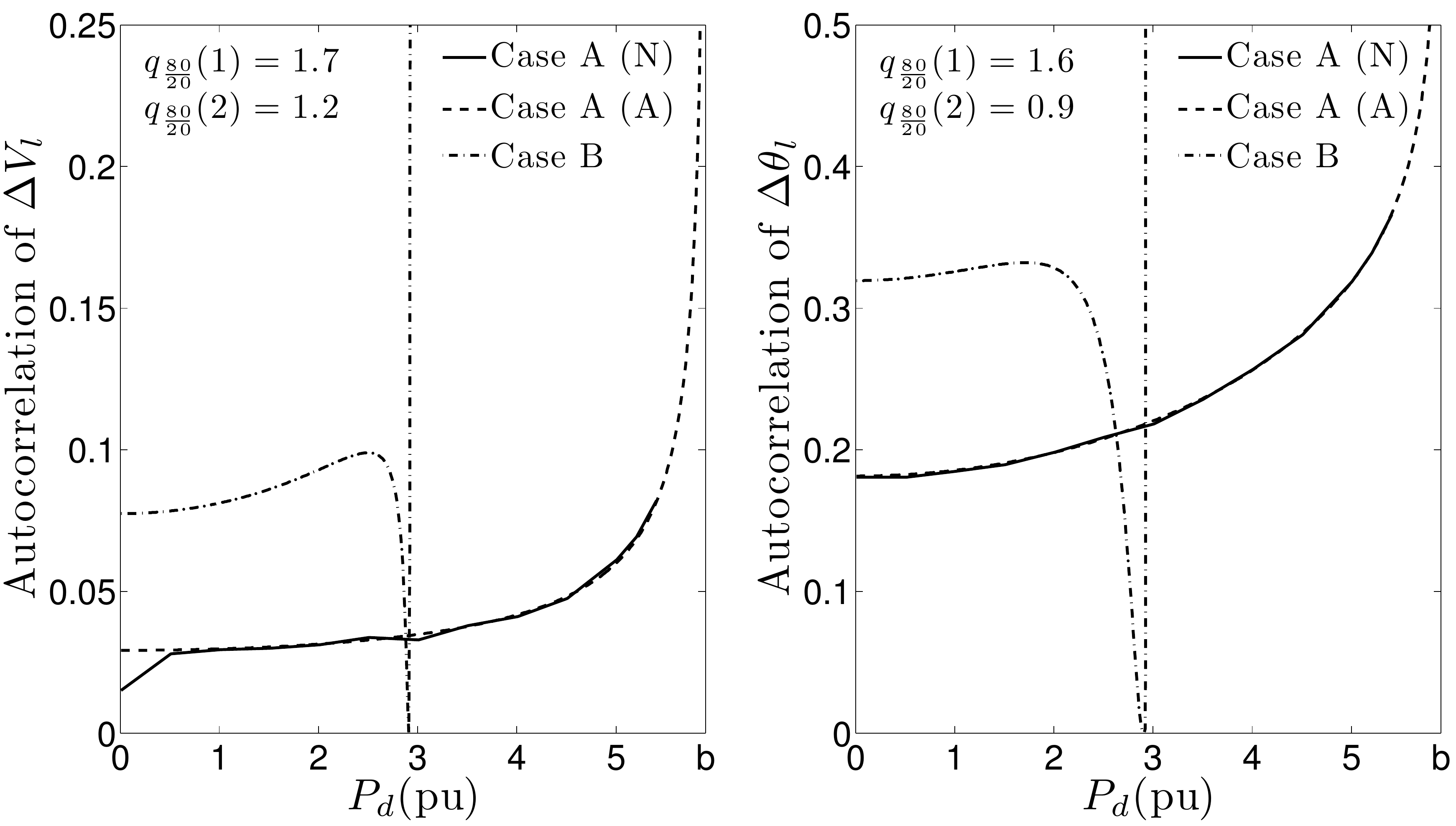}
\par\end{centering}
\vspace{-.15in}
\caption{\label{fig_sys3_ar_vltl}Autocorrelations of $\Delta V_{l},\Delta\theta_{l}$
versus $\textrm{P}_{d}$. }

\vspace{-.2in}
\end{figure}

\par\end{center}

\vspace{-0.95in}

\subsection{Discussion}

After studying this system with a range of different parameters, we
found that  autocorrelations of $ $the differential variables and
variance of the voltage magnitude are consistently good indicators
of  proximity to the bifurcation.

On the other hand, as shown in Fig.~\ref{fig_sys3_var_deom}, variance
in the differential variables is not a reliable indicator. Namely,
variances change non-monotonically (i.e., they do not always increase)
and, importantly, may exhibit very abrupt changes. Fig.~\ref{C6C7}
provides some clues as to the reason for this latter phenomenon. In
this figure, the absolute value of $C_{7}$ decreases with $P_{d}$
and becomes zero very close to the bifurcation point, at $P_{d|C_{7}=0}$.
Therefore, the variances of $\Delta\delta$ and $\Delta\dot{\delta}$,
which are proportional to $C_{7}^{2}$, decrease and vanish at $ $$P_{d|C_{7}=0}$.
Past this point, $\left|C_{7}\right|$ increases, while $C_{6}$ continues
to decrease and vanishes at $b$. Therefore, the variances of $\Delta\delta$
and $\Delta\dot{\delta}$, which are proportional to $\nicefrac{C_{7}^{2}}{C_{6}}$,
increase to infinity in the very narrow interval $\left(P_{d|C_{7}=0},\, b\right)$;
see Fig.~\ref{C6C7}. This explains the sharp features in Fig.~\ref{fig_sys3_var_deom};
a similar explanation can be given to such a feature in Fig.~\ref{fig_sys3_ar_vltl}.
Therefore, neither the variances of $\Delta\delta,\Delta\dot{\delta}$
or the autocorrelations of $\Delta V_{l},\Delta\theta_{l}$ are good
indicators of proximity to bifurcation.

The results for this system clearly show that not all of the variables
in a power system will show CSD signs long before the bifurcation.
Although autocorrelations and variances of all variables increase
before the bifurcation, some of them increase only very near the bifurcation
or the increase is not monotonic.  Hence, these variables are not
useful indicators of proximity to the bifurcation. In the three-bus
system,  autocorrelation in the differential equations was a better
indicator of  proximity than  autocorrelation in $\Delta V_{l}$ or
$\Delta\theta_{l}$, which are not directly associated with the differential
equations. Also, $\Delta V_{l}$ was the only variable whose variance
shows a gradual and monotonic increase with the bifurcation parameter.

\vspace{-.1in}
\section{CSD in multi-machine systems\label{sec:New-England-39-Bus}}

In order to compare these analytical results to results from more
practical power system, this section presents numerical results for
two multi-machine systems. 

The first system was similar to the Three-bus system (case B in Sec.~\ref{sec:Three-Bus-System}).
The only difference was that instead of infinite bus, a generator
similar to the other generator was used. The numerical simulation
results were similar to the Three-bus system, except for the autocorrelation
of $\Delta\dot{\delta}$. Fig.~\ref{fig:2mac_arom} shows that autocorrelation
of $\Delta\dot{\delta}$ increases for one of the machines, while
it decreases for the other one. This shows that the autocorrelation
of $\Delta\dot{\delta}$ is not a reliable indicator of the proximity
to the bifurcation.
\vspace{-.15in}
\begin{figure}[H]
\begin{centering}
\includegraphics[width=1\columnwidth]{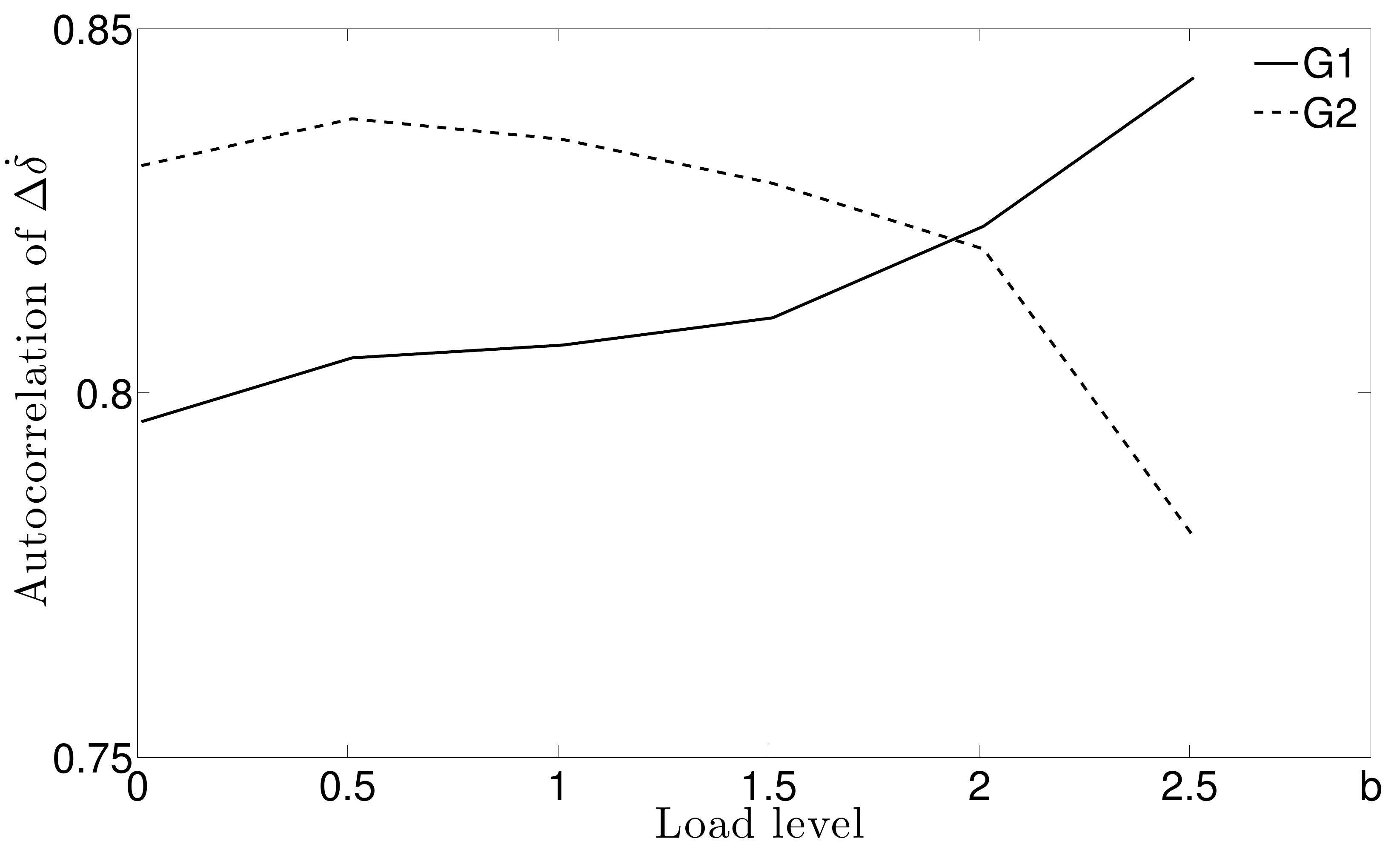}
\par\end{centering}
\vspace{-.15in}
\caption{\label{fig:2mac_arom}Autocorrelation of $\Delta\dot{\delta}$ for
two machines in the Three-bus system with two generators. $G_{1}$
is the same generator as in the Three-bus system and $G_{2}$ is the
new generator.}
\end{figure}
\vspace{-.15in}

The second system we studied was the New England 39-bus system, using
the system data from \cite{pai1989energy} We simulated this system
for different load levels using the power system analysis toolbox
(PSAT) \cite{milano2005open}. Exciters and governors were not included
in the results here, although subsequent tests indicate that adding
them do not substantially change the conclusions. In order to change
the system loading, each load was multiplied by the same factor. At
each load level, we added white noise to each load. As one would expect,
increasing the loads moves the system towards voltage collapse. For
solving the stochastic DAEs, we used the fixed-step trapezoidal solver
of PSAT with the step size of 0.01s. The noise intensity was kept
constant for all load levels.

The simulation results show that the variances and autocorrelations
of bus voltage magnitudes increase with  load. However, similar to
the Three-bus system, the autocorrelations of voltage magnitudes are
very small, indicating that in practice, these variables would not
be good indicators of proximity. The variances and autocorrelations
of generator rotor angles and speeds and bus voltage angles did not
consistently show an  increasing pattern. Figs.~\ref{fig:sys39_V11}
and \ref{fig:sys39_delta5} show the variances and autocorrelations
of the voltage magnitudes of five busbars\textcolor{red}{{} }and the
rotor angles of five generators of the system respectively. The buses
and generators were arbitrarily chosen. As in previous results, the
autocorrelation time lag was chosen to be 0.1s.

The results in this section suggest that autocorrelations of differential
variables show nonmonotic behavior in some cases, which limits their
application as early waning signs of bifurcation.
\vspace{-.1in}

\begin{figure}[H]
\begin{centering}
\includegraphics[width=1\columnwidth]{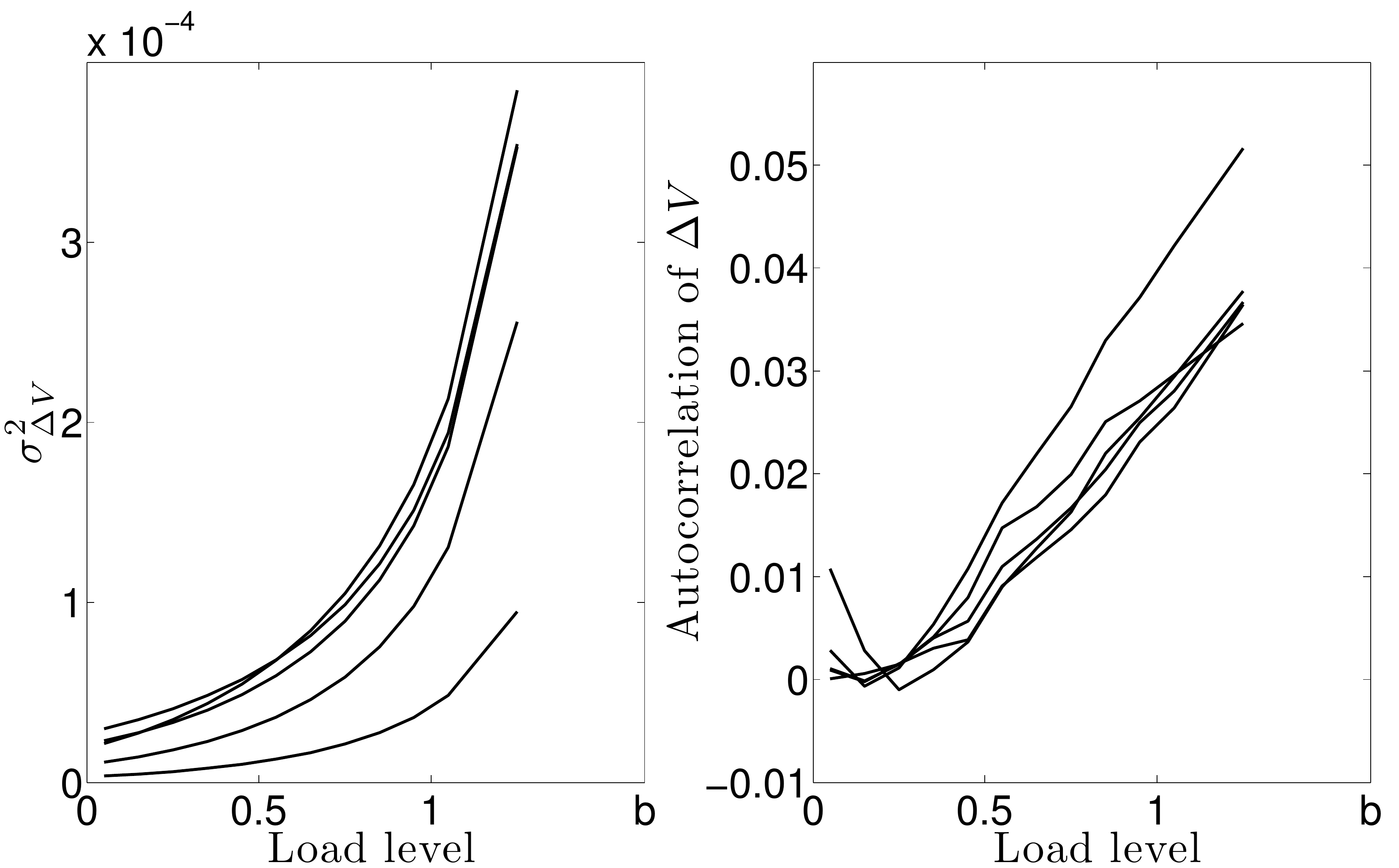}
\par\end{centering}
\vspace{-.1in}

\caption{\label{fig:sys39_V11}The variances and autocorrelations of the voltage
magnitudes of five busbars of the system. Load level is the ratio
of the values of the system's loads to their nominal values.}
\end{figure}
\vspace{-.25in}

\begin{figure}[H]
\begin{centering}
\includegraphics[width=1\columnwidth]{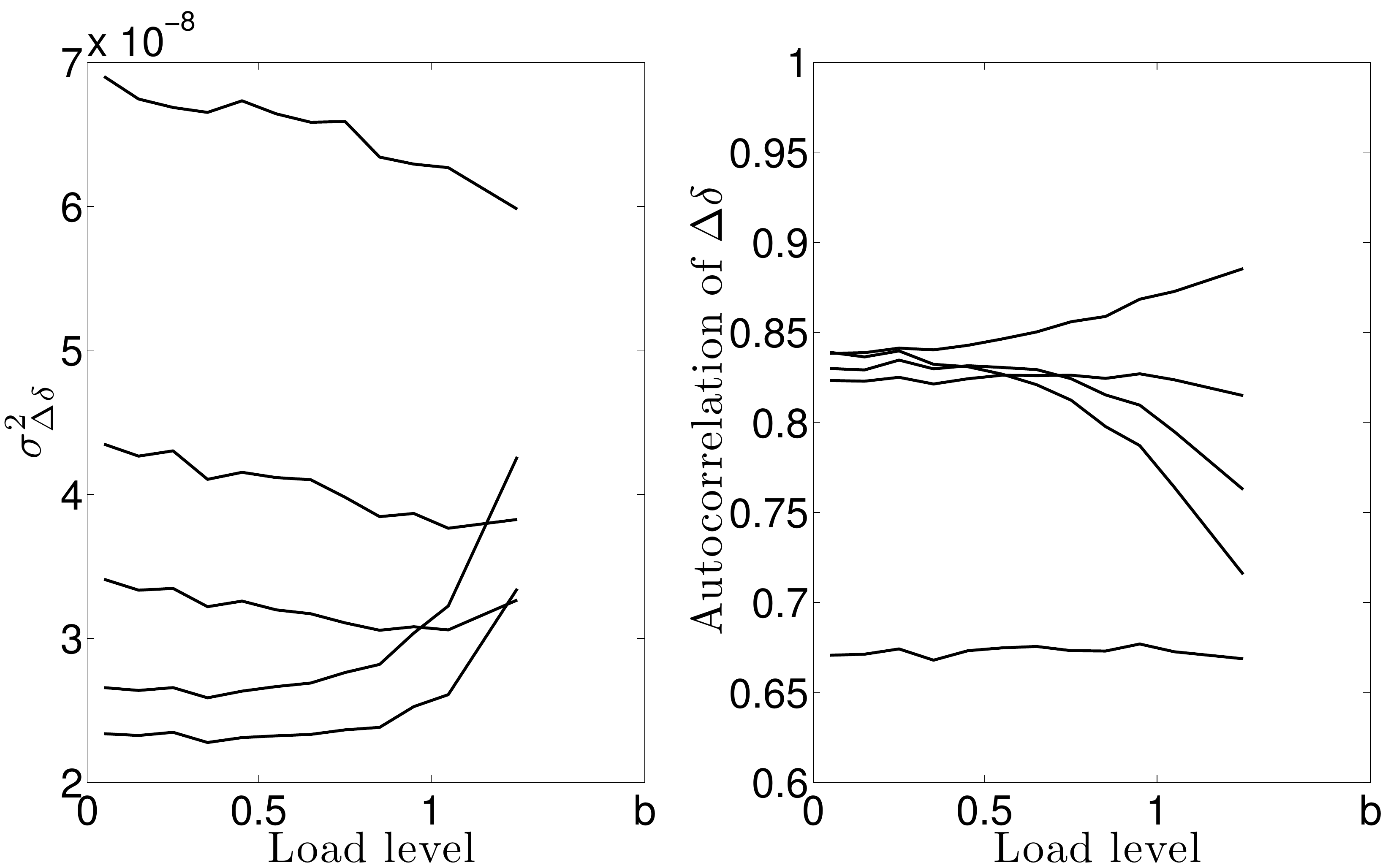}
\par\end{centering}
\vspace{-.15in}

\caption{\label{fig:sys39_delta5}The variances and autocorrelations of the
rotor angles of five generators of the system.}

\vspace{-.15in}
\end{figure}

In many ways, this test case is a multi-machine version of the SMSL
system. As with the SMSL and Three-bus systems, variances of bus voltage
magnitudes are good early warning signs. However, unlike in the SMSL
system, autocorrelation in voltage magnitudes increases, albeit only
slightly, with system load. Unlike in the SMSL system, voltage magnitudes
in the 39-bus case have non-zero autocorrelation for $\Delta t>0$.
This results from the fact that voltage magnitudes are coupled to
the differential variables in this system.

Results from this system, as with the SMSL system, suggest that variance
in voltage magnitudes is a useful early warning sign of voltage collapse.
It is less clear from these results if changes in autocorrelation
will be sufficiently large to provide a reliable early warning of
criticality.

\vspace{-.1in}
\section{Conclusion\label{sec:Conclusion}}

In this paper, we analytically and numerically solve the stochastic differential algebraic equations
 for three small power system models in order to understand critical slowing down in power systems.
 The results from the single machine infinite bus system and the Three-bus system models show that
 critical slowing down does occur in power systems, and illustrate that autocorrelation and variance in
 some cases can be good indicators of proximity to criticality in power systems. The results also show
 how non-linear dynamics influence the observed changes in autocorrelation and variance. For example,
 linearity of the differential equation in the single machine single load system caused the autocorrelation
 of the differential variable to be constant. On the other hand, in the SMIB system and Three-bus system,
 the differential equations were nonlinear and autocorrelations of the differential variables increased with
 the bifurcation parameter.

Although the signs of critical slowing down do consistently appear as the systems approach bifurcation,
 only in a few of the variables did the increases in autocorrelation appear sufficiently early to give a useful
 early warning of potential collapse. On the other hand, variance in load bus voltages consistently showed
 substantial increases with load, indicating that variance in bus voltages can be a good indicator of voltage
 collapse in multi-machine power system models. This was verified for the New England 39-bus system. 

Together these results suggest that it is possible to obtain useful information about system stability
 from high-sample rate time-series data, such as that produced by synchronized phasor measurement units.
 Future research will focus on developing an effective power system stability indicator based on these results. 
\vspace{-.1in}

\appendices

\section{\label{appendix:SMSL} }

The derivation of (\ref{eq:35-7}) is presented in this section. By
linearizing (\ref{eq:27}) around the equilibrium and replacing the
obtained equation for $P_{g}$ in (\ref{eq:26}), we obtained the
following:
\begin{equation}
M\Delta\ddot{\delta}+D\Delta\dot{\delta}=-C_{12}\Delta V_{l}-C_{13}\left(\Delta\delta-\Delta\theta_{l}\right)\label{eq:30}
\end{equation}
where $C_{12}$ and $C_{13}$ are:
\begin{eqnarray}
C_{12} & = & E_{a}^{'}\sin\left(\theta_{l0}-\delta_{0}-\arctan\left(\frac{G_{gl}}{B_{gl}}\right)\right)\label{eq:30-2}\\
 &  & \cdot\sqrt{G_{gl}^{2}+B_{gl}^{2}}\nonumber \\
C_{13} & = & V_{l0}E_{a}^{'}\cos\left(\theta_{l0}-\delta_{0}-\arctan\left(\frac{G_{gl}}{B_{gl}}\right)\right)\label{eq:30-3}\\
 &  & \cdot\sqrt{G_{gl}^{2}+B_{gl}^{2}}\nonumber 
\end{eqnarray}
By linearizing (\ref{eq:31}) and (\ref{eq:32}) around the equilibrium,
and solving for $\Delta V_{l}$ and $\Delta\delta-\Delta\theta_{l}$,
we obtained the following:
\begin{eqnarray}
\Delta V_{l} & = & C_{14}\eta\label{eq:33}\\
\Delta\delta-\Delta\theta_{l} & = & C_{15}\eta\label{eq:34}
\end{eqnarray}
where $C_{14}$ and $C_{15}$ are:
\begin{eqnarray}
C_{14} & = & \frac{C_{19}P_{d0}-C_{17}Q_{d0}}{C_{17}C_{18}-C_{16}C_{19}}\label{eq:35-1}\\
C_{15} & = & \frac{C_{18}P_{d0}-C_{16}Q_{d0}}{C_{17}C_{18}-C_{16}C_{19}}\label{eq:35-2}
\end{eqnarray}
where $C_{16}-C_{19}$ are given below:
\begin{eqnarray}
C_{16} & = & E_{a}^{'}\sin\left(\theta_{l0}-\delta_{0}+\arctan\left(\frac{G_{gl}}{B_{gl}}\right)\right)\label{eq:35-3}\\
 &  & \cdot\sqrt{G_{gl}^{2}+B_{gl}^{2}}+2G_{ll}V_{l0}\nonumber \\
C_{17} & = & V_{l0}E_{a}^{'}\cos\left(\theta_{l0}-\delta_{0}+\arctan\left(\frac{G_{gl}}{B_{gl}}\right)\right)\label{eq:35-4}\\
 &  & \cdot\sqrt{G_{gl}^{2}+B_{gl}^{2}}\nonumber \\
C_{18} & = & -E_{a}^{'}\cos\left(\theta_{l0}-\delta_{0}+\arctan\left(\frac{G_{gl}}{B_{gl}}\right)\right)\label{eq:35-5}\\
 &  & \cdot\sqrt{G_{gl}^{2}+B_{gl}^{2}}-2B_{ll}V_{l0}\nonumber \\
C_{19} & = & V_{l0}E_{a}^{'}\sin\left(\theta_{l0}-\delta_{0}+\arctan\left(\frac{G_{gl}}{B_{gl}}\right)\right)\label{eq:35-6}\\
 &  & \cdot\sqrt{G_{gl}^{2}+B_{gl}^{2}}\nonumber 
\end{eqnarray}
Using (\ref{eq:33}) and (\ref{eq:34}), we rewrote (\ref{eq:30})
as (\ref{eq:35-7}) where $C_{5}$ is as follows:
\begin{equation}
C_{5}=\frac{\left(C_{13}C_{18}+C_{12}C_{19}\right)P_{d0}-\left(C_{13}C_{16}+C_{12}C_{17}\right)Q_{d0}}{C_{16}C_{19}-C_{17}C_{18}}\label{eq:35-8}
\end{equation}

\section{\label{appendix:sys3_1}}

The derivation of $C_{6},C_{7}$ is presented in this section. By
using (\ref{eq:2-1}) and linearizing (\ref{eq:39})-(\ref{eq:41})
around the equilibrium, we have the following:
\begin{eqnarray}
\Delta\ddot{\delta} & = & -\left(D\Delta\dot{\delta}+C_{20}\Delta V_{l}+C_{21}\left(\Delta\delta-\Delta\theta_{l}\right)\right)/M\label{eq:43}\\
0 & = & -P_{d0}\eta+C_{22}\Delta V_{l}+C_{21}\Delta\delta+C_{23}\Delta\theta_{l}\label{eq:44}\\
0 & = & -\Delta V_{l}+C_{24}\Delta\delta+C_{25}\Delta\theta_{l}\label{eq:45}
\end{eqnarray}
where $C_{20}$ through $C_{25}$ are as follows:
\begin{eqnarray}
C_{20}\left(\delta_{0},\theta_{l0}\right) & = & \frac{E'_{a}}{X}\sin\left(\delta_{0}-\theta_{l0}\right)\label{eq:46}\\
C_{21}\left(\delta_{0},\theta_{l0},V_{l0}\right) & = & \frac{E'_{a}V_{l0}}{X}\cos\left(\delta_{0}-\theta_{l0}\right)\label{eq:47}\\
C_{22}\left(\delta_{0},\theta_{l0}\right) & = & C_{20}\left(\delta_{0},\theta_{l0}\right)-\frac{\sin\left(\theta_{l0}\right)}{X_{l2}}\label{eq:48}\\
C_{23}\left(\delta_{0},\theta_{l0},V_{l0}\right) & = & -C_{21}\left(\delta_{0},\theta_{l0},V_{l0}\right)\label{eq:49}\\
 &  & -\frac{V_{l0}}{X_{l2}}\cos\left(\theta_{l0}\right)\nonumber \\
C_{24}\left(\delta_{0},\theta_{l0}\right) & = & -\beta E'_{a}\sin\left(\delta_{0}-\theta_{l0}\right)\label{eq:50}\\
C_{25}\left(\delta_{0},\theta_{l0}\right) & = & -C_{24}\left(\delta_{0},\theta_{l0}\right)-\left(1-\beta\right)\nonumber \\
 &  & \cdot\sin\left(\theta_{l0}\right)\label{eq:51}
\end{eqnarray}
where $\beta=X_{l2}/(X+X_{l2})$. Using (\ref{eq:44}) and (\ref{eq:45}),
we solved for $\Delta V_{l}$ and $\Delta\theta_{l}$:
\begin{eqnarray}
\Delta V_{l} & = & C_{26}\Delta\delta+C_{27}\eta\label{eq:52}\\
\Delta\theta_{l} & = & C_{28}\Delta\delta+C_{29}\eta\label{eq:53}
\end{eqnarray}
where $C_{26}$ through $C_{29}$ are as follows:
\begin{eqnarray}
C_{26} & = & \frac{C_{23}C_{24}-C_{21}C_{25}}{C_{22}C_{25}+C_{23}}\label{eq:54}\\
C_{27} & = & \frac{C_{25}P_{d0}}{C_{22}C_{25}+C_{23}}\label{eq:55}\\
C_{28} & = & -\frac{C_{21}+C_{22}C_{24}}{C_{22}C_{25}+C_{23}}\label{eq:56}\\
C_{29} & = & \frac{P_{d0}}{C_{22}C_{25}+C_{23}}\label{eq:57}
\end{eqnarray}
Equations (\ref{eq:43}), (\ref{eq:52}) - (\ref{eq:57}) lead to
the following expressions for $C_{6}$ and $C_{7}$:
\begin{eqnarray}
C_{6} & = & C_{21}C_{28}-C_{20}C_{26}-C_{21}\label{eq:59}\\
C_{7} & = & C_{21}C_{29}-C_{20}C_{27}\label{eq:60}
\end{eqnarray}

\section*{Acknowledgment}
The authors acknowledge Christopher Danforth for helpful contributions to this research, as well as
 the Vermont Advanced Computing Core, which is supported by NASA (NNX-08AO96G), for providing
 computational resources. 

\renewcommand\]{\end{equation}}

\bibliographystyle{IEEEtran}
\nocite{*}
\bibliography{CAS_bib}

\section*{Author biographies}
\vspace{-.4in}
\begin{IEEEbiographynophoto}{Goodarz Ghanavati} (S`11) received the B.S. and M.S. degrees in Electrical Engineering from Amirkabir University of Technology, Tehran, Iran in 2005 and 2008, respectively. Currently, he is pursuing the Ph.D. degree in Electrical Engineering at University of Vermont. His research interests include power system dynamics, PMU applications and smart grid.
\end{IEEEbiographynophoto}
\vspace{-.4in}

\begin{IEEEbiographynophoto}{Paul D.~H.~Hines} (S`96,M`07) received the Ph.D. in Engineering and Public Policy from Carnegie Mellon University in 2007 and M.S. (2001) and B.S. (1997) degrees in Electrical Engineering from the University of Washington and Seattle Pacific University, respectively.\\
He is currently an Assistant Professor in the School of Engineering, and the Dept. of Computer Science at the University of Vermont, and a member of the adjunct research faculty at the Carnegie Mellon Electricity Industry Center. Formerly he worked at the U.S. National Energy Technology Laboratory, the US Federal Energy Regulatory Commission, Alstom ESCA, and Black and Veatch. He currently serves as the vice-chair of the IEEE Working Group on Understanding, Prediction, Mitigation and Restoration of Cascading Failures, and as an Associate Editor for the IEEE Transactions on Smart Grid. He is National Science Foundation CAREER award winner.
\end{IEEEbiographynophoto}
\vspace{-.4in}
\vfill

\begin{IEEEbiographynophoto}{Taras I. Lakoba} received the Diploma in physics from Moscow State University, Moscow, Russia, in 1989, and the Ph.D. degree in applied mathematics from Clarkson University, Potsdam, NY, in 1996.\\
In 2000 he joined the Optical Networking Group at Lucent Technologies, where he was engaged in the development of an ultralong-haul terrestrial fiber-optic transmission system. Since 2003 he has been with the Department of Mathematics and Statistics of the University of Vermont. His research interests include multichannel all-optical regeneration, the effect of noise in fiber-optic communication systems, stability of numerical methods for nonlinear wave equations, and perturbation techniques.
\end{IEEEbiographynophoto}
\vspace{-.4in}

\begin{IEEEbiographynophoto}{Eduardo Cotilla-Sanchez} (S`08,M`12) received the M.S. and Ph.D. degrees in electrical engineering from the University of Vermont, Burlington, in 2009 and 2012, respectively. He is currently an Assistant Professor in the School of Electrical Engineering and Computer Science at Oregon State University, Corvallis. His primary field of research is electrical infrastructure protection, in particular, the study of cascading outages. Cotilla-Sanchez is a member of the IEEE Cascading Failure Working Group.  
\end{IEEEbiographynophoto}

\vfill
\end{document}